\documentclass[aps,twocolumn,pra,footinbib,longbibliography]{revtex4-1}
\usepackage{amsmath,amssymb,bm}
\usepackage{graphicx}
\usepackage{epstopdf}
\usepackage{latexsym}
\usepackage{subfigure}
\usepackage[usenames,dvipsnames]{color}
\usepackage{hyperref}
\usepackage{natbib}
\usepackage{moreverb}
\begin{document}

\newcommand{\ns}{\mathcal{N}_{\mathrm{s}}}
\newcommand{\nb}{\mathcal{N}_{\mathrm{b}}}
\newcommand{\warn}[1]{{\color{red}\textbf{* #1 *}}}

\newcommand{\eqplan}[1]{{\color{blue}\textbf{equations:{ #1 }}}}

\newcommand{\figplan}[1]{{\color{blue}\textbf{figures: { #1 }}}}

\newcommand{\tableplan}[1]{{\color{blue}\textbf{tables: { #1 }}}}

\newcommand{\warntoedit}[1]{{\color{blue}\textbf{EDIT: #1 }}}

\newcommand{\warncite}[1]{{\color{green}\textbf{cite #1}}}

\title{Simulating generic spin-boson models with matrix product states}
\date{\today}
\author{Michael L. Wall}
\affiliation{JILA, NIST and University of Colorado, 440 UCB, Boulder, CO 80309, USA}
\affiliation{Center for Theory of Quantum Matter, University of Colorado, Boulder, Colorado 80309, USA}
\author{ Arghavan Safavi-Naini}
\affiliation{JILA, NIST and University of Colorado, 440 UCB, Boulder, CO 80309, USA}
\affiliation{Center for Theory of Quantum Matter, University of Colorado, Boulder, Colorado 80309, USA}
\author{Ana Maria Rey}
\affiliation{JILA, NIST and University of Colorado, 440 UCB, Boulder, CO 80309, USA}
\affiliation{Center for Theory of Quantum Matter, University of Colorado, Boulder, Colorado 80309, USA}

\begin{abstract}

The global coupling of few-level quantum systems (``spins") to a discrete set of bosonic modes is a key ingredient for many applications in quantum science, including large-scale entanglement generation, quantum simulation of the dynamics of long-range interacting spin models, and hybrid platforms for force and spin sensing.  We present a general numerical framework for treating the out-of-equilibrium dynamics of such models based on matrix product states.  Our approach applies for generic spin-boson systems: it treats any spatial and operator dependence of the two-body spin-boson coupling and places no restrictions on relative energy scales.  We show that the full counting statistics of collective spin measurements and infidelity of quantum simulation due to spin-boson entanglement, both of which are difficult to obtain by other techniques, are readily calculable in our approach.  We benchmark our method using a recently developed exact solution for a particular spin-boson coupling relevant to trapped ion quantum simulators.  Finally, we show how decoherence can be incorporated within our framework using the method of quantum trajectories, and study the dynamics of an open-system spin-boson model with spatially non-uniform spin-boson coupling relevant for trapped atomic ion crystals in the presence of molecular ion impurities.

\end{abstract}

\maketitle

\section{Introduction}

The coupling of spins to bosonic modes provides a paradigmatic setting for various important phenomena in many-body physics, such as decoherence, thermalization, virtually mediated long-range spin-spin interactions, and large-scale entanglement generation~\cite{PhysRevLett.104.073602,bohnet2014reduced,KasevichSqueezing}.   The study of spin-boson models has historically belonged to the domain of condensed matter theory~\cite{caldeira1983quantum}.  However, experimental setups available to atomic, molecular and optical (AMO) physics provide versatile, clean, and controllable realizations of such models. In particular, spin-boson models can be engineered in  trapped ion systems by coupling the discrete phonon modes of the ion crystal to the ion spins~\cite{PhysRevLett.82.1971,PhysRevLett.92.207901,PhysRevLett.103.120502,britton2012engineered,PhysRevA.78.010101} and in cavities via cold atoms~\cite{RevModPhys.85.553,PhysRevLett.104.073602} or artificial atoms built from superconducting qubits~\cite{niemczyk2010circuit,hur2015many} coupled to quantized cavity modes; similar models can also be achieved in optomechanical setups~\cite{RevModPhys.86.1391}.  While these AMO platforms differ from the most common condensed matter spin-boson analog--a two-state version of the Caldeira-Leggett model of a particle interacting with a continuum of harmonic oscillators--due to their discrete boson mode spectra, they offer clean systems in which one can observe out-of-equilibrium dynamics.

The long history of spin-boson models has led to the development of a range of theoretical approaches.   The most prevalent one is to adiabatically eliminate the bosons, keeping only their virtual effect on the spins in the form of long-range spin-spin couplings~\cite{PhysRevLett.92.207901}.  While this approach is useful in contexts where there is a very large separation of energy scales between the bosons and spins, in many experimental realizations and applications such an approach is invalid.  An additional powerful approach, valid in the case that only uniform, collective spin operators appear in the Hamiltonian, uses the permutational symmetry of the density matrix to reduce the computational scaling of direct numerical diagonalization from exponential to polynomial in the number of spins~\cite{Carmichael,hartmann2012generalized, PhysRevB.91.035306}.  Outside of the uniform coupling regime, analytical results have recently been obtained in a special case in which the coupling Hamiltonian is linear in boson operators and commutes with the spin Hamiltonian~\cite{Dylewsky}.  However, in spite of the importance of the many-spin, many boson problem to myriad fields of physics, a general, systematic approach to study their quantum dynamics has yet to be developed.

\begin{figure}[t]
\centering
\includegraphics[width=0.9\columnwidth]{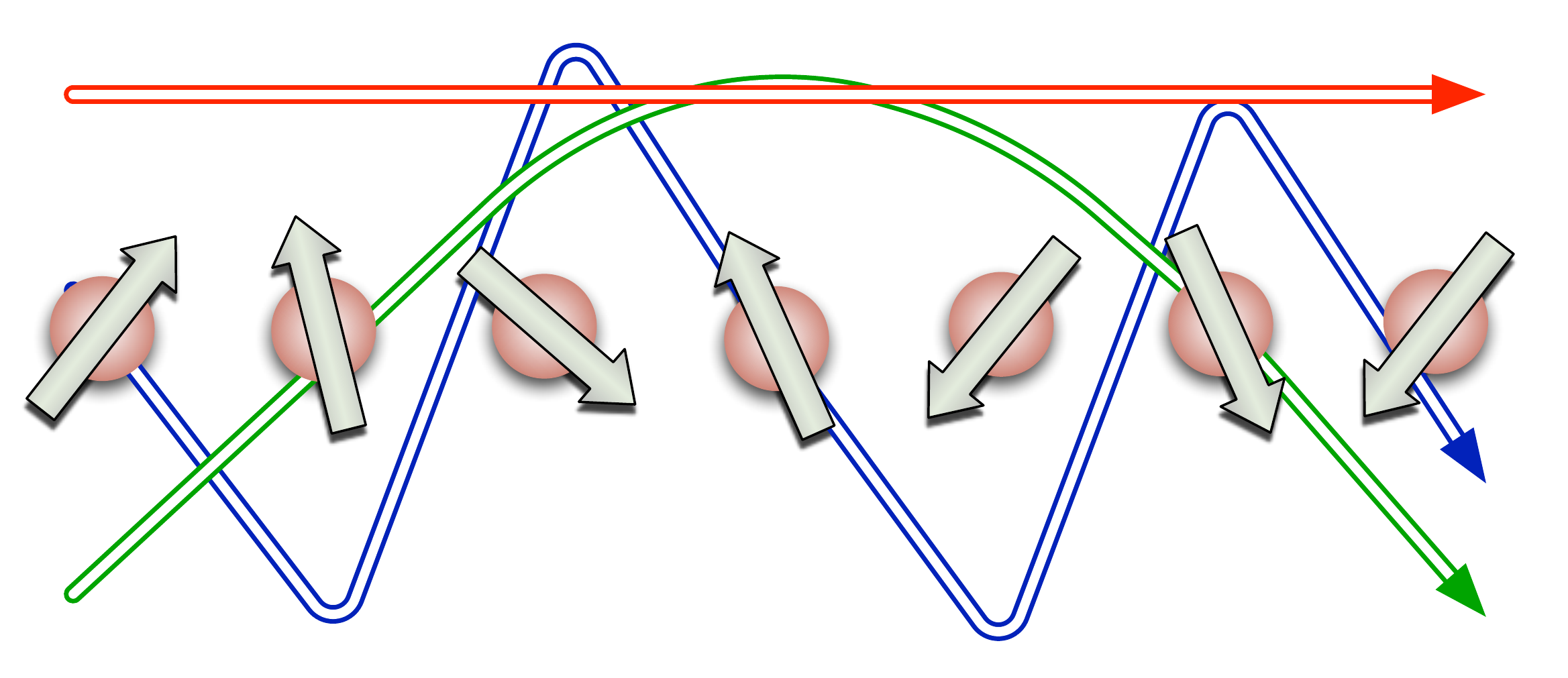}
\caption{(Color online) \emph{Spins globally coupled to bosonic modes.}  A schematic description of $\mathcal{N}_s=7$ spins coupled globally to $\mathcal{N}_b=3$ bosonic modes (colored lines).  Position in the array corresponds to, e.g., position of the spins in real space.  The amplitude of the boson line denotes the value of the spin-boson coupling at that position.  Our approach applies for any spatial dependence of the bosonic mode coupling.
}
\label{fig:Snake!}
\end{figure}

In what follows, we show that spin-boson models with global coupling can be systematically treated with well-characterized error using the framework of matrix product states (MPSs).  Our methodology consists of two complementary methods for treating the out-of-equilibrium dynamics.  The first, using the technology of swap gates~\cite{TTN} to deform the MPS topology dynamically, is most efficient in the case of few bosons or at weak coupling, but has a rigorously bounded, systematically correctable error.  The second uses an exact matrix product operator (MPO) representation of the spin-boson Hamiltonian, and time-evolves under this Hamiltonian using recently developed techniques for time evolution with long-range couplings~\cite{Zaletel}, together with safeguards against variational metastability.  This latter method is most efficient when the number of bosonic modes is equal to the number of spins, relevant for trapped ion quantum simulators in certain cases.  

This paper is organized as follows: Sec.~\ref{sec:Form} presents our numerical formalism, which includes a brief review of matrix product states (MPSs) and operators as well as our new MPS-based methodologies for spin-boson systems; Sec.~\ref{sec:Benchmark} discusses benchmark calculations of our methods for 1D and 2D systems of trapped ions in the parameter regimes of recent experiments~\cite{richerme2014non,bohnet2015quantum}; Sec.~\ref{sec:QT} extends our formalism to open quantum systems through the technique of quantum trajectories; and Sec.~\ref{sec:Impure} presents simulations of trapped ion systems in the presence of impurity ions where decoherence and spatially non-uniform spin-boson couplings invalidate many other approaches; Finally, in Sec.~\ref{sec:Concl} we conclude and give an outlook on broad applications of our methods.  Some technical results are given as appendices.

\section{Formalism}
\label{sec:Form}

We consider a collection of $\ns$ spins localized on a lattice and coupled globally to $\nb$ bosonic modes, shown schematically in Fig.~\ref{fig:Snake!}.  We take the Hamiltonian to have the generic form
\begin{align}
\label{eq:HFull}\hat{H}&=\hat{H}_{\mathrm{spin}}+\hat{H}_{\mathrm{boson}}+\hat{H}_{\mathrm{s-b}}\, ,
\end{align}
where $\hat{H}_{\mathrm{spin}}$ acts only on the spins, $\hat{H}_{\mathrm{boson}}$ only on the bosons, and the spin-boson coupling has the form
\begin{align}
\hat{H}_{\mathrm{s-b}}&=\sum_{\mu=1}^{\nb}\sum_{i=1}^{\ns}\hat{H}_{\mathrm{s-b};\mu,i}\\
\label{eq:HSB}&=\sum_{\mu=1}^{\nb}\sum_{i=1}^{\ns}\sum_{\alpha=1}^{n}g^{\left(\alpha\right)}_{\mu i}\hat{X}^{\left(\alpha\right)}_{\mu}\hat{Y}^{\left(\alpha\right)}_{i}\, .
\end{align}
Here and throughout, roman letters $i$ refer to spin indices and greek letters $\mu$ to boson indices.  The operators $\hat{X}^{\left(\alpha\right)}_{\mu}$ and $\hat{Y}^{\left(\alpha\right)}_{i}$ can be taken to be Hermitian, and act on bosons and spins, respectively.  The index $\alpha$ counts the number of such pairs of Hermitian operators, e.g., $(\hat{a}_{\mu}^{\dagger}+\hat{a}_{\mu})\hat{\sigma}^z_j$ has $n=1$ while $(\hat{a}^{\dagger}_{\mu}\hat{\sigma}^-_j+\hat{a}_{\mu}\hat{\sigma}^+_j)=(1/2)(\hat{a}_{\mu}^{\dagger}+\hat{a})\sigma^x_j-(i/2)(\hat{a}_{\mu}^{\dagger}-\hat{a})\sigma^y_j$ has $n=2$.  While we will continually refer to our degrees of freedom as being spins, our approach can be straightforwardly generalized from spins to mobile particles with any quantum statistics.  A key feature of this general model is that the coupling between bosons and spins is arbitrary, i.e.~each boson mode $\mu$ can be coupled to each spin $i$ with an arbitrary amplitude and operator pairs $\{\hat{X}^{\left(\alpha\right)}_{\mu},\hat{Y}^{\left(\alpha\right)}_{i}\}$.  In contrast to approaches based on permutational symmetry of the density matrix~\cite{Carmichael,hartmann2012generalized, PhysRevB.91.035306}, we do not require that this coupling is uniform; our approach applies for any amplitudes $g^{\left(\alpha\right)}_{\mu i}$ and operators $\hat{X}^{\left(\alpha\right)}_{\mu}$ and $\hat{Y}^{\left(\alpha\right)}_{i}$.  Many iconic models can be cast in the form Eq.~\eqref{eq:HFull}, including the Rabi~\cite{gardas2011exact,Braak}, Jaynes-Cummings~\cite{JC,Eberly}, and Tavis-Cummings~\cite{Agarwal} models of quantum optics and optomechanics~\cite{RevModPhys.86.1391} and the Holstein-Hubbard~\cite{Jeckelmann} model of condensed matter physics.  Also, we point out that our setting is more general than previous spin-boson approaches using related density-matrix renormalization group (DMRG) methods, in which each spin was coupled locally to a single boson~\cite{Jeckelmann,Bursill,Brockt}, a single spin was coupled to many bosons~\cite{Guo,Frenzel,schroder2015simulating}, or a single collective boson mode was coupled to many spins~\cite{Gammelmark}.

\subsection{Matrix product states and operators}

Before proceeding to the details of how we simulate the out-of-equilibrium dynamics of Eq.~\eqref{eq:HFull}, we remind the reader a few basic facts about matrix product states (MPSs) and matrix product operators (MPOs); a more detailed discussion can be found in, e.g.,~\cite{Schollwoeck,Orus}.  The natural setting of MPSs is a 1D chain with $L$ sites, where lattice site $i$ can be in one of the $d_i$ states $|j_i\rangle$.  An MPS is a representation of a many-body quantum state on such a lattice, defined as
\begin{align}
\label{eq:MPSdef}|\psi\rangle&=\sum_{j_1\dots j_L}\mathrm{Tr}\left[A^{\left[1\right]j_1}\dots A^{\left[L\right]j_L}\right]|j_1\rangle\otimes\dots \otimes |j_L\rangle\, .
\end{align}
Here, $A^{\left[i\right]j_i}$ is a $\chi_{i-1}\times \chi_i$ dimensional matrix, and the maximum value of any of the $\chi_k$ is called the bond dimension $\chi$ of the MPS.  MPSs are exact ground states of 1D gapped models with short-range interactions~\cite{hastings2007area}.  In addition, MPSs have proven to be a useful variational ansatz for long-range interacting systems~\cite{PhysRevB.78.035116,PhysRevX.3.031015, Wall_Carr_12}, gapless systems, e.g., quantum critical points~\cite{Pollmann,rams2014truncating}, and quasi-higher-dimensional systems~\cite{Mishmash, Stoudenmire_2D,Depenbrock} by ``snaking" a 1D line across the sites of the higher-dimensional lattice.

\begin{figure}
\centering
\includegraphics[width=0.7\columnwidth]{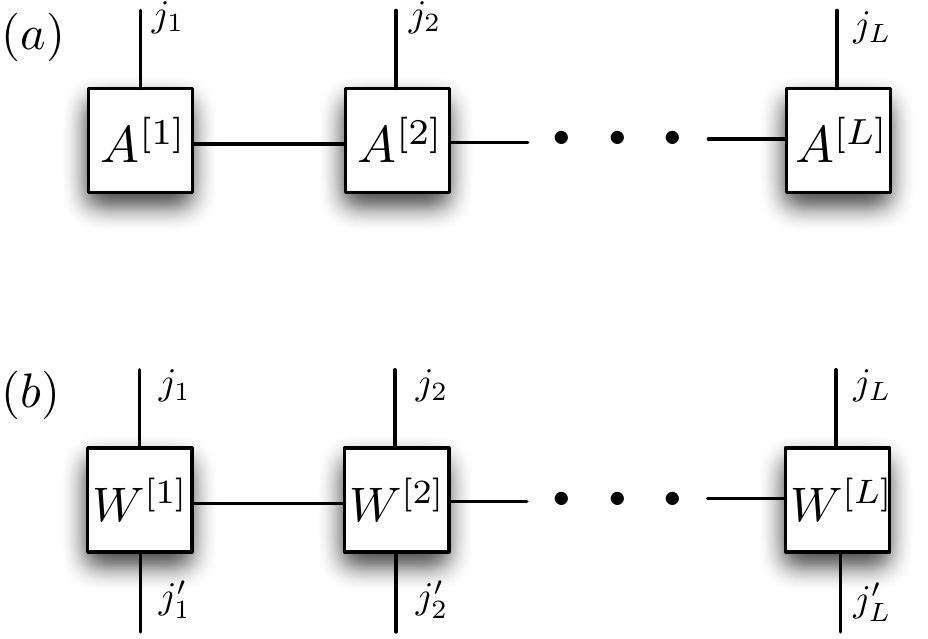}
\caption{(a) Tensor network diagram representation of an MPS with open boundary conditions.  Lines extending to the top of the page correspond to physical states $|j_i\rangle$ of local system $i$.  (b) Tensor network diagram representation of an MPO with open boundary conditions.  Each tensor has two physical indices, corresponding to a linear transformation (operator).
}
\label{fig:TND}
\end{figure}
A matrix product operator (MPO) is the natural operator-valued extension of the MPS definition~\cite{mcculloch2007density,PhysRevLett.93.207204}.  It is a representation of a many-body operator acting on the same space defined for the MPS above, and has the form
\begin{align}
\nonumber \hat{O}&=\sum_{j_1,j_1'\dots j_L,j_L'}\mathrm{Tr}\left[W^{\left[1\right]j_1,j_1'}\dots W^{\left[L\right]j_L,j_L'}\right]\\
\label{eq:MPOdef1}&\times |j_1\rangle\langle j_1'|\otimes\dots \otimes |j_L\rangle\langle j_L'|\, .
\end{align}
MPOs are of interest because (1) they take MPSs to MPSs (with, in general, larger bond dimension), (2) most many-body operators of interest, e.g. Hamiltonians, can be represented as MPOs with constant bond dimension by using a small set of MPO construction ``rules"~\cite{PhysRevA.78.012356,PhysRevA.81.062337,pirvu2010matrix,Wall_Carr_12}, and (3) many MPS algorithms, such as variational search for the ground state and time evolution, can be formulated generically given that all many-body operators of interest have a known MPO form~\cite{Wall_Carr_12}.  It is useful to re-write Eq.~\eqref{eq:MPOdef1} as
\begin{align}
\label{eq:MPOdef2}\hat{O}&=\mathrm{Tr}\left[\mathcal{W}^{\left[1\right]}\otimes \dots \otimes \mathcal{W}^{\left[L\right]}\right]\, ,
\end{align}
where each one of the $\mathcal{W}^{\left[i\right]}$ is an operator-valued matrix of the form, e.~g.,
\begin{align}
\mathcal{W}^{\left[i\right]}&=\left(\begin{array}{cc} \hat{A}&\hat{B}\\ \hat{C}&\hat{D}\end{array}\right)\, .
\end{align}
The matrix indices are the fictitious ``bond" indices contracted in the trace of Eq.~\eqref{eq:MPOdef2}, and the indices of the individual operators, e.g.,~$\hat{A}$, run over the $d_i$ states of the local physical Hilbert space.  For simplicity, we have given a bond dimension two MPO matrix as an example, but any dimensionality is possible (see Eq.~\eqref{eq:HMPO}).  A useful means of displaying MPSs, MPOs, and their operations is the Penrose tensor network diagram notation exemplified in Fig.~\ref{fig:TND}.  In this notation, a tensor is an object with lines extending from it, with the number of lines equal to the rank of the tensor.  A line connecting two tensors denotes that index is summed, or ``contracted," between the two tensors.  

\subsection{Swap operator decomposition of the spin-boson state and propagator}
\label{sec:MI}

In this section, we propose a method for simulating the out-of-equilibrium dynamics of Eq.~\eqref{eq:HFull} which is based on Trotter-Suzuki decompositions.  First, we use a Trotter-Suzuki decomposition to isolate the action of $\hat{H}_{\mathrm{s-b}}$ from $\hat{H}_{\mathrm{spin}}$ and $\hat{H}_{\mathrm{boson}}$.  The simplest non-trivial such decomposition is ($\hbar=1$ unless otherwise specified throughout)
\begin{align}
e^{-i \delta t\hat{H} }&=e^{-i \frac{\delta t}{2}\left(\hat{H}_{\mathrm{spin}}+\hat{H}_{\mathrm{boson}}\right)}e^{-i\delta t \hat{H}_{\mathrm{s-b}}}e^{-i \frac{\delta t}{2}\left(\hat{H}_{\mathrm{spin}}+\hat{H}_{\mathrm{boson}}\right)}\, ,
\end{align}
which is valid to order $\mathcal{O}\left(\delta t^3\right)$.  While this is written in a form appropriate for time-independent $\hat{H}$, a time-ordered version of this formula also exists~\cite{Huyghebaert_DeRaedt}, as well as higher order decompositions in $\delta t$~\cite{omelyan2002optimized}.  Since $\hat{H}_{\mathrm{spin}}$ and $\hat{H}_{\mathrm{boson}}$ commute by construction, the evolutions associated with these operators can be performed in any order.  In the examples given in this paper both $\hat{H}_{\mathrm{spin}}$ and $\hat{H}_{\mathrm{boson}}$ are sums of single-site operators and so their exponentials are product operators which are trivial to apply to an MPS.  Alternatively, in the most commonly encountered case that they are short-ranged, this propagation can be performed with ordinary t-DMRG~\cite{PhysRevLett.93.040502,daley2004time}.

We now turn to the application of $\exp(-i\delta t \hat{H}_{\mathrm{s-b}})$ to an MPS.  We do so by using another Trotter-Suzuki decomposition to write this exponential as a product of two-site unitaries, each acting on a single spin and boson pair.  The simplest such decomposition is
\begin{align}
\nonumber e^{-i\delta t\hat{H}_{\mathrm{s-b}}}&=\left[\prod_{i=\mathcal{N}_s}^{1}\left(\prod_{\mu=\mathcal{N}_b}^{1} e^{-\frac{i\delta t}{2}\hat{H}_{\mathrm{s-b};\mu,i}}\right)\right]\\
\label{eq:STsb}&\times \left[\prod_{i=1}^{\mathcal{N}_s}\left(\prod_{\mu=1}^{\mathcal{N}_b} e^{-\frac{i\delta t}{2}\hat{H}_{\mathrm{s-b};\mu,i}}\right)\right]+\mathcal{O}\left(\delta t^3\right)\, ,
\end{align}
and, as before, higher-order decompositions can be devised~\cite{Sornborger_Stewart}.  In this expression, $\prod_{\mu=\mathcal{N}_b}^{1}\hat{O}_{\mu}$ means the product $\hat{O}_{\mathcal{N}_b}\hat{O}_{\mathcal{N}_b-1}\dots \hat{O}_{1}$.  For a fixed ordering of MPS sites, the application of Eq.~\eqref{eq:STsb} requires coupling of sites which are not contiguous.  While this can be done with a variety of available long-range time evolution methods for MPSs~\cite{Manmana,GarciaRipoll,Wall_Carr_12,Zaletel,haegeman2014unifying}, such an approach can suffer from numerical issues, e.g. getting stuck far from the variational optimum.  In contrast, applying an operator which couples neighboring sites in the 1D MPS representation can be performed efficiently, without issues of variational metastability, and with a precisely controlled error~\cite{PhysRevLett.93.040502}.

We can apply the decomposed operator Eq.~\eqref{eq:STsb} to an MPS using only operations on neighboring sites by inserting swap gates $\hat{S}_{\mu i}$ which interchange the positions of boson $\mu$ and spin $i$ in the MPS representation between different terms appearing in the product~\cite{TTN}.  Swap gates have been previously used to implement MPS algorithms for periodic boundary conditions~\cite{Ippei}, as well as for generic long-range time evolution~\cite{Stoudenmire_Metts}.  In the latter case, long-range time evolution requires $\mathcal{O}\left(L^2\right)$ swaps per time step, where $L$ is the total number of lattice sites; this can be prohibitively costly.  In our case, only $\mathcal{O}\left(\mathcal{N}_s\mathcal{N}_b\right)$ swaps are required, which is more favorable in the often realized case that $\mathcal{N}_s\gg \mathcal{N}_b$ as well as the case $\mathcal{N}_b\gg \mathcal{N}_s$.

\begin{figure}[t]
\centering
\includegraphics[width=0.75\columnwidth]{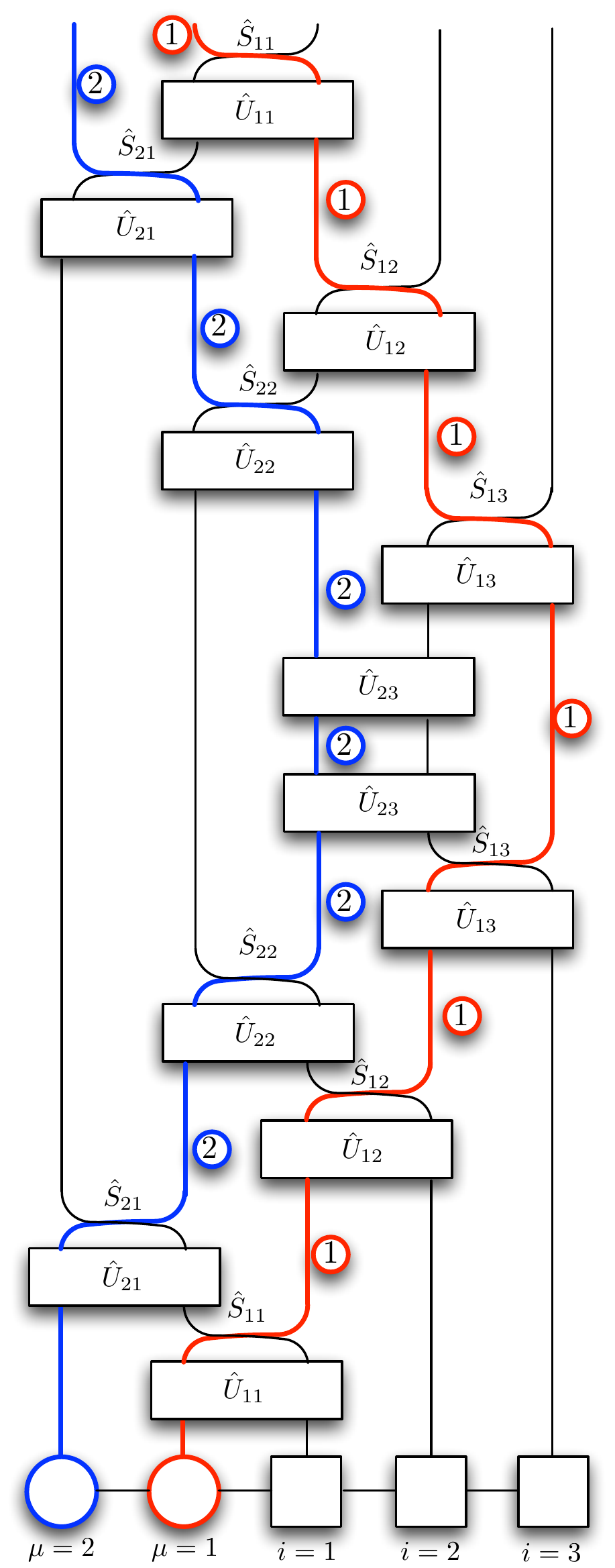}
\caption{(Color online) \emph{Application of spin-boson propagator through swap gates.}  The tensor network diagram for the application of $e^{-i\delta t\hat{H}_{\mathrm{s-b}}}$ on a state of $\mathcal{N}_b=2$ bosons (circles) and $\mathcal{N}_s=3$ spins (squares) through the Trotter-Suzuki decomposition with swap gates (crossed lines) demonstrates that only nearest-neighbor operations are required.  For clarity, circles containing the boson mode number are drawn next to boson index lines, and the boson lines have been color-coded.
}
\label{fig:Swapping}
\end{figure}

To see how the swapping works in practice, we show the tensor network diagram for a full application of $e^{-i\delta t\hat{H}_{\mathrm{s-b}}}$ for $\mathcal{N}_s=3$ and $\mathcal{N}_b=2$ in Fig.~\ref{fig:Swapping}.  Here, the tensors in the initial MPS are circles (squares) for boson (spin) sites, $\hat{U}_{\mu i}=\exp(-i\delta t \hat{H}_{\mathrm{s-b};\mu,i})$, and crossed lines indicate the application of a swap gate.  Motivated by the form of the decomposition Eq.~\eqref{eq:STsb}, a natural 1D chain ordering of the spins and bosons is to begin with all of the bosons left of the spins.  In order to keep track of the position of the boson sites in the MPS representation, we draw a circle containing the mode index next to index lines corresponding to bosons, showing that each boson is moved past all of the spins using the swap gates.  From the diagram it is clear that the combination of Trotter-Suzuki decomposition and swap gates leads to a propagator which involves only operations on nearest-neighbor sites.

\subsection{Exact MPO representation of the spin-boson Hamiltonian}

As the number of bosons approaches the number of spins, the scaling of the swapping method described in the last section approaches $\mathcal{O}(\ns^2)$, which can be prohibitively expensive.  However, as we will discuss below, in this situation a direct simulation of the dynamics of the spin-boson system using a representation of the Hamiltonian with long-range couplings between spins and bosons can be efficient.  Here, the ordering of the spins and bosons in the MPS representation can be chosen to minimize issues of variational metastability that can pose problems for long-range time evolution with MPSs.  Before we discuss how to choose this ordering, we first show that Eq.~\eqref{eq:HFull} has an exact, compact MPO representation for spin-boson couplings of the form Eq.~\eqref{eq:HSB}, irrespective of how the spins and bosons are ordered in the 1D array.  For simplicity, we will write down the MPO in the case that $\hat{H}_{\mathrm{spin}}=\sum_{i=1}^{\mathcal{N}_s}\hat{H}_{s;i}$ and $\hat{H}_{\mathrm{boson}}=\sum_{\mu=1}^{\mathcal{N}_b}\hat{H}_{b;\mu}$ do not couple multiple sites and the sum over $\alpha$ in Eq.~\eqref{eq:HSB} contains a single term; this is the form of the Hamiltonian relevant for trapped ion experiments, as shown in Sec.~\ref{sec:Benchmark}.  More complex spin and boson Hamiltonians can be formulated using known MPO constructions in the literature~\cite{PhysRevA.78.012356,PhysRevA.81.062337,pirvu2010matrix,Wall_Carr_12}, and more terms in the $\alpha$ summation can be included by matrix direct sums~\cite{mcculloch2007density}.  Additionally, for compactness, we consider the case of two boson modes; the MPO for many modes is also a straightforward generalization.  

To begin, we write the collection of $\mathcal{N}_s$ spins and $2$ bosons as a 1D chain with $(\mathcal{N}_s+2)$ sites, and let the boson modes $\mu=1$ and 2 be placed at sites $j_1$ and $j_2$, respectively, with $j_1<j_2$.  Then, the MPO matrices representing Eq.~\eqref{eq:HFull} take the form
\begin{align}
\label{eq:HMPO}\mathcal{W}^{\left[j< j_1\right]}&=\left(\begin{array}{cccc} \hat{I}_s&0&0&0\\ 0&\hat{I}_s&0&0\\ 0&0&\hat{I}_s&0\\ \hat{H}_{s;i_j}&g_{1i_j}^{\left(1\right)}\hat{Y}^{\left(1\right)}_{i_j}&g_{2i_j}^{\left(1\right)}\hat{Y}^{\left(1\right)}_{i_j}&\hat{I}_s\end{array}\right)\, ,\\
\nonumber \mathcal{W}^{\left[j_1\right]}&=\left(\begin{array}{cccc} \hat{I}_b&0&0&0\\ \hat{X}^{\left(1\right)}_1 &0&0&0\\ 0 &0&\hat{I}_b&0\\ \hat{H}_{b;1}& \hat{X}^{\left(1\right)}_1&0&\hat{I}_b\end{array}\right)\, , \\
\nonumber \mathcal{W}^{\left[j_1<j< j_2\right]}&=\left(\begin{array}{cccc} \hat{I}_s&0&0&0\\ g_{1i_j}^{\left(1\right)}\hat{Y}^{\left(1\right)}_{i_j}&\hat{I}_s&0&0\\ 0&0&\hat{I}_s&0\\ \hat{H}_{s;i_j}&0&g_{2i_j}^{\left(1\right)}\hat{Y}^{\left(1\right)}_{i_j}&\hat{I}_s\end{array}\right)\, ,\\
\nonumber \mathcal{W}^{\left[j_2\right]}&=\left(\begin{array}{cccc} \hat{I}_b&0&0&0\\ 0 &\hat{I}_b&0&0\\ \hat{X}^{\left(1\right)}_2 &0&0&0\\ \hat{H}_{b;2}&0& \hat{X}^{\left(1\right)}_2&\hat{I}_b\end{array}\right)\, , 
\end{align}
\begin{align}
\nonumber \mathcal{W}^{\left[j> j_2\right]}&=\left(\begin{array}{cccc} \hat{I}_s&0&0&0\\ g_{1i}^{\left(1\right)}\hat{Y}^{\left(1\right)}_{i_j}&\hat{I}_s&0&0\\ g_{2i_j}^{\left(1\right)}\hat{Y}^{\left(1\right)}_{i_j}&0&\hat{I}_s&0\\ \hat{H}_{s;i_j}&0&0&\hat{I}_s\end{array}\right)\, .
\end{align}
Here, $\hat{I}_s$ ($\hat{I}_b$) is the identity operator in the spin (boson) space, $i_j$ is the spin index for lattice site $j$, and the first (last) MPO matrix for open boundary conditions is the last row (first column) of the associated MPO matrix $\mathcal{W}$.  Remarkably, in spite of the generality and complexity of the model, Eq.~\eqref{eq:HFull}, the Hamiltonian admits a very simple and compact representation.  This MPO is useful not only for the out-of-equilibrium dynamics we discuss in the next section, but can also be used to find eigenstates variationally~\cite{Schollwoeck}.

\subsection{Long-range time evolution with $\mathcal{N}_s=\mathcal{N}_b$}
\label{sec:MII}

Many methods for time evolution of MPSs exist which are applicable to general Hamiltonians in MPO form, such as the Krylov~\cite{Manmana,GarciaRipoll,Wall_Carr_12} and the local Runge-Kutta~\cite{Zaletel} methods, which require some form of variational optimization of an MPS, and the time-dependent variational principle~\cite{haegeman2014unifying}, which has an additional error associated with projection of the Hamiltonian onto a local subspace.  While such methods are useful in many cases, the error analysis and convergence behavior of these algorithms can be complex.  Of particular concern when applying these algorithms to our Hamiltonian \eqref{eq:HFull} via the exact MPO representation Eq.~\eqref{eq:HMPO} is that the effectiveness will depend strongly on the ordering of the spin and boson modes in the 1D MPS representation.  This is especially true when many strongly coupled boson modes are present, as this situation can lead to glassy spin physics~\cite{gopalakrishnan2009emergent,grass2015controlled} which may pose problems for variational optimization.  

\begin{figure}[t]
\centering
\includegraphics[width=0.75\columnwidth]{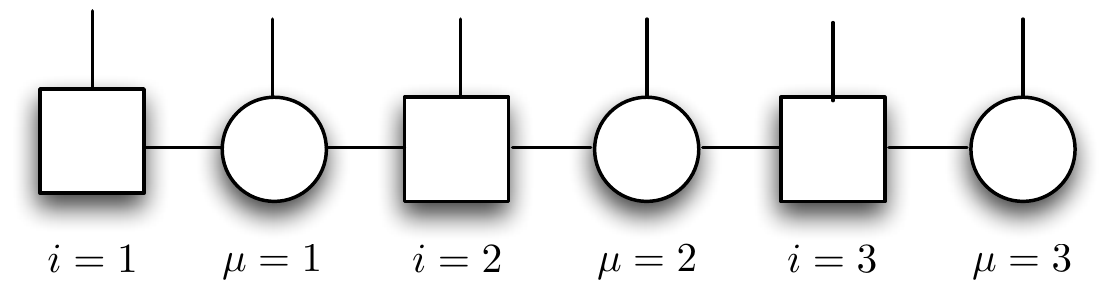}
\caption{\emph{MPS topology for the long-range time evolution method with $\mathcal{N}_s=\mathcal{N}_b=3$.}  To avoid variational metastability when directly simulating time-evolution with long-ranged spin-boson interactions, we use an MPS representation in which spins are always neighbored by bosons, allowing for strong spin-boson entanglement to be built up locally.  The coherence between spins and bosons which are not neighboring is built up both by repeated variational sweeping and by mixing in a small amount of a state which contains long-range spin-boson coherence, see the text for details.
}
\label{fig:IIdia}
\end{figure}

The main issue of variational metastability for systems with long-range interactions is that efficient MPS algorithms use an iterative local variational optimization to search for the global optimum~\cite{Schollwoeck}, where local means that two neighboring sites are optimized at a time.  Hence, MPS algorithms build up entanglement and correlations locally, two sites at a time, and then build up longer-ranged correlations and entanglement by repeated ``sweeping," in which each pair of neighboring sites are optimized in a round-robin fashion.  If there are long-ranged (farther than nearest-neighbor) terms in the Hamiltonian, entanglement resulting from these terms is not directly built in by the optimization, and so we must either build in the proper entanglement structure by hand or it must be incorporated in by other short-range terms in the Hamiltonian upon repeated sweeping.

For the system at hand and in the case $\mathcal{N}_s=\mathcal{N}_b$, we can ensure that all spins and bosons have a least some short-range entanglement built in during the optimization by alternating the spins and bosons in the 1D representation, see Fig.~\ref{fig:IIdia}.  In this representation, each spin $i$ neighbors boson $\mu=(i-1)$ and $\mu=i$, and hence entanglement between this spin and these bosons is directly generated as part of a two-site variational optimization procedure.  The fluctuations of these boson modes can then couple to the fluctuations of further away spins, improving convergence to the global optimum.  A further safeguard against variational metastability, in which a state with long-range coherence between spins and bosons far-separated in the MPS representation is explicitly mixed in during optimization to speed convergence, is described in Appendix~\ref{sec:VM}.  Our method of choice for long-range time evolution is the second-order local Runge-Kutta method of Ref.~\cite{Zaletel}.

\subsection{Observables and fidelities}
\label{sec:OaF}
\begin{figure}
\centering
\includegraphics[width=0.75\columnwidth]{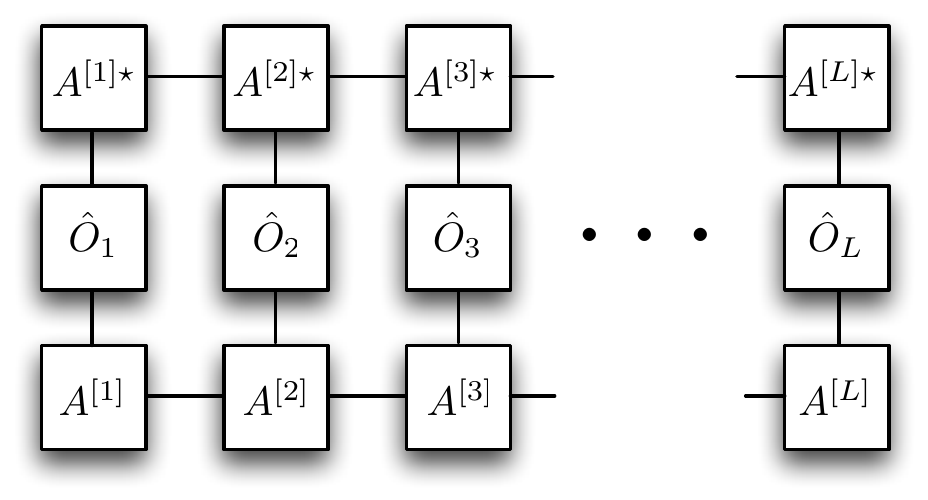}
\caption{\emph{Tensor network representation of the characteristic function.} The tensor network description of the characteristic function for a given Fourier variable $k$ and spin quadrature $\mathbf{n}$ is especially simple due to the fact it is the expectation of a product operator.  Here, $\hat{O}_j=\exp[{\pi i k\hat{\boldsymbol{\sigma}}_j\cdot \mathbf{n} /(\ns+1)}]$.
}
\label{fig:FCS}
\end{figure}

While the computation of observables with MPSs, e.g. single-spin expectations or two-point spin-spin correlation functions, is standard and discussed in review articles~\cite{Schollwoeck,Orus}, here we point out the measurement of two quantities which are non-standard.  The first is the measurement of the full probability distribution of outcomes for a collective spin measurement along direction $\mathbf{n}$, i.e. the full counting statistics (FCS) $P_m\left(\mathbf{n}\right)$ for $m$ spins to be aligned along $\mathbf{n}$.  To compute the FCS directly involves the computation of all correlation functions of any order, which in general scales poorly with the number of spins~\cite{Dylewsky}.  However, the FCS can be computed efficiently given an MPS representation for the state of the spins by measuring the Fourier transform of this distribution, known as the characteristic function
\begin{align}
C_k\left(\mathbf{n}\right)&=\langle e^{\pi i k \sum_{j=1}^{\ns}\hat{\boldsymbol{\sigma}}_j\cdot \mathbf{n} /(\ns+1)}\rangle \, ,
\end{align}
$k=0,\dots,\ns$, and then inverse Fourier transforming.  Noting that
\begin{align}
C_k\left(\mathbf{n}\right)&=\langle \prod_{j=1}^{\ns}e^{\pi i k\hat{\boldsymbol{\sigma}}_j\cdot \mathbf{n} /(\ns+1)}\rangle  \, ,
\end{align}
the characteristic function for each $k$ is the expectation of an MPO with bond dimension 1, which can be performed at the same $\mathcal{O}\left(L\chi^3\right)$ cost as taking the overlap of two MPSs, see Fig.~\ref{fig:FCS}.  The FCS is useful in trapped ion quantum simulators because it is readily accessible thanks to near single-ion resolution, and it provides detailed information about the structure of the state beyond low-order correlation functions~\cite{bohnet2015quantum}.

\begin{figure}
\centering
\includegraphics[width=0.85\columnwidth]{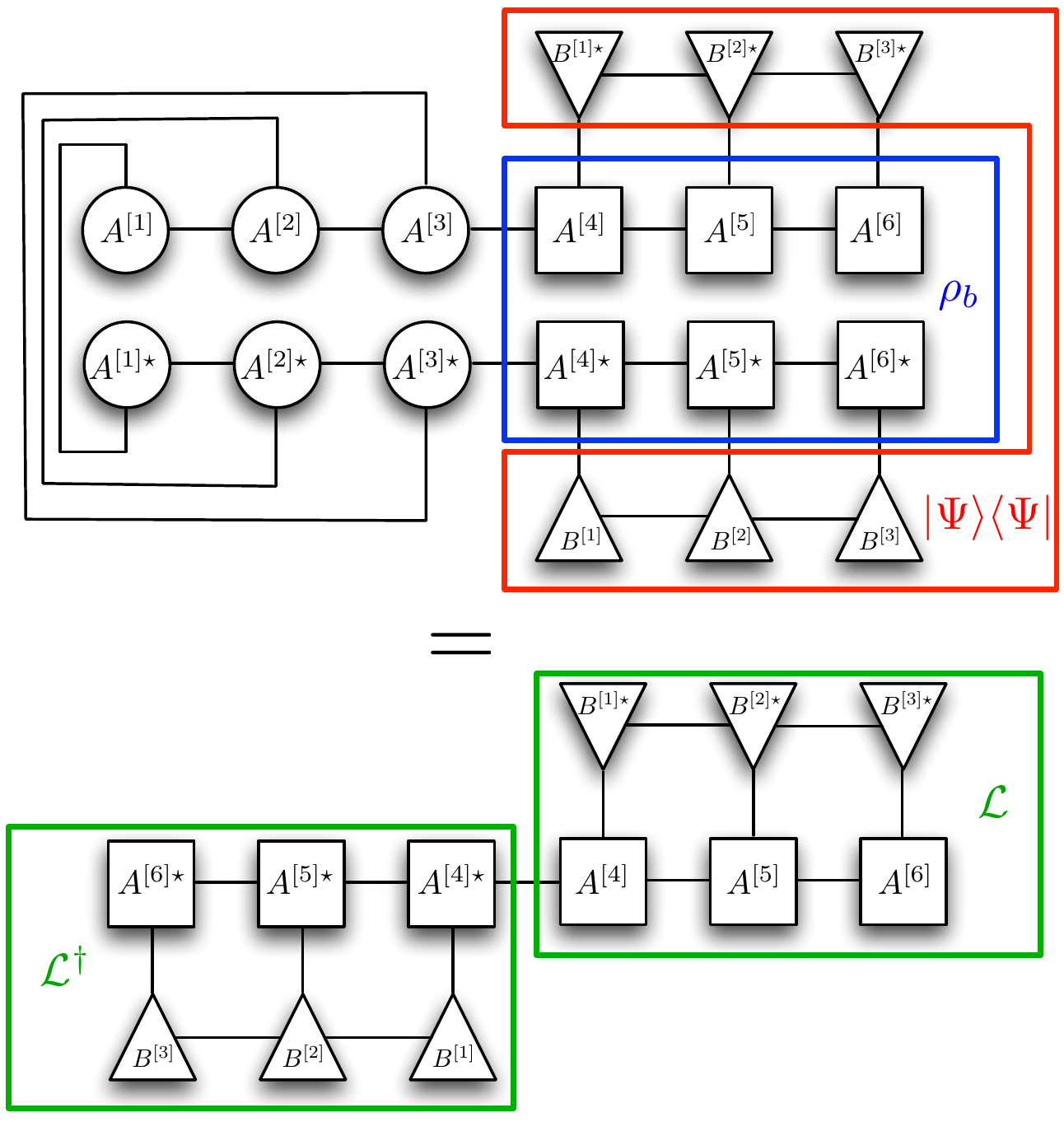}
\caption{(Color online) \emph{Tensor network representation of the fidelity between a traced spin-boson state and a pure spin state.} An example tensor network diagram giving the (squared) fidelity of the spin density matrix $\rho_b$ of a system with three bosons (circles) and three spins (squares) with the bosons traced out with the pure density matrix $|\Phi\rangle\langle \Phi|$ of a system comprised of only spins (triangles).  Assuming the boson tensors are in left canonical form, the upper diagram is transformed to the lower diagram, which is the inner product of a vector $\mathcal{L}$.
}
\label{fig:Fidelity}
\end{figure}

Often, rather than simulating a complex model of spins coupled to bosons, one would like to integrate out the bosons and keep only their virtual effect in the form of an effective model acting only on the spins.  The most stringent test of the faithfulness of this spin-only description is the fidelity between the reduced density matrix of the spin-boson system obtained by tracing out the bosons with that of a pure spin state evolving under an effective spin Hamiltonian.  This fidelity is defined as
\begin{align}
F\left(\rho_{b},\rho\right)&=\mathrm{Tr}\left(\sqrt{\sqrt{\rho}\rho_b \sqrt{\rho}}\right)\, ,
\end{align}
where $\rho_{b}=\mathrm{Tr}_{\mu}|\psi\rangle\langle \psi|$ is the density matrix obtained from tracing the bosons out from the pure spin-boson state $|\psi\rangle$ and $\rho=|\Psi\rangle\langle\Psi|$ is the density matrix of pure state $|\Psi\rangle$ containing only spins.  As $\rho$ is pure, we find that
\begin{align}
F\left(\rho_{b},\rho\right)&=\sqrt{\langle \Psi|\rho_b|\Psi\rangle}\, .
\end{align}
As we intend to trace out the bosons, it is useful to separate the bosons and spins spatially in the MPS representation.  Such a representation naturally appears in our swapping dynamics protocol, see Fig.~\ref{fig:Swapping}, and can be obtained straightforwardly from any representation using swap gates.  With this ordering, the fidelity has a simple tensor network representation as the norm of a vector $\mathcal{L}$, provided both $|\psi\rangle$ and $|\Psi\rangle$ have known MPS representations and the boson tensors are in left-canonical form~\cite{Schollwoeck}, as shown in Fig.~\ref{fig:Fidelity}.  The restriction on canonical form, which involves no loss of generality, simply uses the freedom inherent in the MPS ansatz to transform the contraction over boson states into the identity, making the trace over bosonic degrees of freedom particularly simple.  

\section{Benchmark calculations for 1D and 2D trapped ion systems}
\label{sec:Benchmark}

As an application of the above methods, we will consider crystals of trapped ions subject to a spin-dependent force.  Such systems have been a topic of great recent interest due to their capability of performing high-fidelity quantum simulation of long-range interacting quantum spin models~\cite{PhysRevLett.92.207901}.  In these systems, a competition between the Coulomb repulsion of the ions and external trapping from electromagnetic potentials leads to a crystalline structure for the ions in equilibrium.  The normal modes of these equilibrium crystal structures comprise a set of global boson (phonon) modes, which are then coupled to the spin of the ions using lasers in a Raman scheme with characteristic wavevector $k_{R}$ to impart a spin-dependent force along the direction perpendicular to the crystal structure~\cite{leibfried2003experimental,milburn2000ion,PhysRevA.62.022311,PhysRevLett.82.1971,islam2013emergence}.  After performing a frame transformation on the spins, the Hamiltonian is expanded to lowest order in the Lamb-Dicke parameters $\eta_{\mu}=k_R\sqrt{\hbar^2/2M\omega_{\mu}}$, where $M$ is the mass of an ion and $\{\omega_{\mu}\}$ are the phonon normal mode frequencies.  In addition, we will transform the phonons to a frame rotating with $\omega_R$, the Raman beatnote frequency of the beams creating the spin-dependent force, and enact a rotating wave approximation.  While this approximation will fail for very large detunings, we have verified that it is an excellent approximation for the parameter regimes explored in this work, and it leads to time-independent Hamiltonians which are more amenable to numerical computation.  Following these steps, the Hamiltonian takes the form of Eq.~\eqref{eq:HFull} with $\hat{H}_{\mathrm{spin}}=0$, $\hat{H}_{\mathrm{boson}}=-\sum_{\mu=1}^{\nb}\delta_{\mu} \hat{n}_{\mu}$, and
\begin{align}
 \label{eq:Hions}\hat{H}_{\mathrm{s-b}}&=-\frac{1}{2}\sum_{j=1}^{\ns}\sum_{\mu=1}^{\nb} \Omega_{\mu} b_{j\mu}(\hat{a}_{\mu}+\hat{a}_{\mu}^{\dagger})\hat{\sigma}^z_j\, .
\end{align}
Here, $\delta_{\mu}=\omega_R-\omega_{\mu}$ is the detuning of mode $\mu$ from the Raman beatnote frequency, $\Omega_{\mu}=F \eta_{\mu}/k_R$, with $F$ the magnitude of the spin-dependent force, quantifies the strength of the spin-dependent force for mode $\mu$, and $\mathbf{b}_{\mu}$ is the normal mode amplitude vector of phonon $\mu$.

The effective Hamiltonian describing the time-dependent dynamics of Eq.~\eqref{eq:Hions} can be found by moving to an interaction picture which rotates with $\hat{H}_{\mathrm{boson}}$ and using the Magnus series~\cite{Magnus_54,Blanes_Casas_09}.  Due to the fact that all spin operators in $\hat{H}_{\mathrm{s-b}}$ commute and the boson operators form $\nb$ copies of the Heisenberg algebra, the Magnus series truncates exactly at second order, and the propagator may be written exactly within the rotating wave approximation as
\begin{align}
\hat{U}_I\left(t\right)&=\hat{U}_{\mathrm{SP}}\left(t\right)\hat{U}_{\mathrm{SS}}\left(t\right)\, ,
\end{align}
with the spin-phonon coupling propagator
\begin{align}
\hat{U}_{\mathrm{SP}}\left(t\right)&= \exp[\sum_{\mu} \sum_j\left(\alpha_{\mu j}\left(t\right)\hat{a}_{\mu}^{\dagger}-\bar{\alpha}_{\mu j}\left(t\right)\hat{a}_{\mu}\right)\hat{\sigma}^z_j]
\end{align}
and the spin-spin coupling propagator
\begin{align}
\hat{U}_{\mathrm{SS}}\left(t\right)&=\exp[-i \sum_{i,j}\tilde{J}_{ij}\left(t\right)\hat{\sigma}^z_i\hat{\sigma}^z_j]\, .
\end{align}
In these expressions, we have defined
\begin{align}
\label{eq:alphadef} \alpha_{\mu j}(t)&=\Omega_{\mu} b_{\mu j}\left(\frac{e^{-i\delta_{\mu} t}-1}{2\delta_{\mu}}\right)\, ,
\end{align}
and 
\begin{align}
\label{eq:SpinSpinTD}\tilde{J}_{ij}\left(t\right)&=\frac{1}{4}\sum_{\mu} \Omega^2_{\mu}\frac{b_{\mu i}b_{\mu j}}{\delta_{\mu}^2}\left(\delta_{\mu} t-\sin\delta t\right)
\end{align}
Noting that the last term in the braces of Eq.~\eqref{eq:SpinSpinTD} is bounded while the first grows without bound, $\tilde{J}_{jj'}\left(t\right)$ is commonly approximated as $\tilde{J}_{ij}\left(t\right)\approx J_{ij} t$, where
\begin{align}
\label{eq:SSapprx}J_{ij}&=\sum_{\mu} \frac{\Omega^2_{\mu} b_{\mu i}b_{\mu j}}{4\delta_{\mu}}\, .
\end{align}
In this approximation, $\hat{U}_{\mathrm{SS}}\left(t\right)$ is the propagator of a long-range Ising model $\hat{H}_{\mathrm{Ising}}=\sum_{i,j}J_{i,j}\hat{\sigma}^z_i\hat{\sigma}^z_j$. 

Trapped ion quantum spin simulator experiments can be divided into two regimes parameterized by their hierarchy of energy scales.  This categorization also currently coincides with the effective dimensionality of the ion crystal in present experiments, with 1D crystals corresponding to linear Paul traps~\cite{PhysRevLett.103.120502}, and 2D systems forming in Penning traps~\cite{britton2012engineered}.  Experiments in 1D Paul traps typically operate in the regime $\delta_{\mu}\gg \Omega_{\mu}$, in which all modes of the ion crystal participate in the dynamics.  In contrast, current experiments in the 2D Penning trap operate in the regime where $\delta_{\mu}\gg \Omega_{\mu}$ for all modes except for the center of mass mode $\tilde{\mu}$, for which $\delta_{\tilde{\mu}}\gtrsim \Omega_{\tilde{\mu}}$.  Hence, the center of mass mode dominates the dynamics in these experiments.  We stress that both types of traps can operate in both parameters regimes in principle, and our grouping of parameter regimes with dimensionality refers only to current experiments.

\begin{figure}
\centering
\includegraphics[width=0.99\columnwidth]{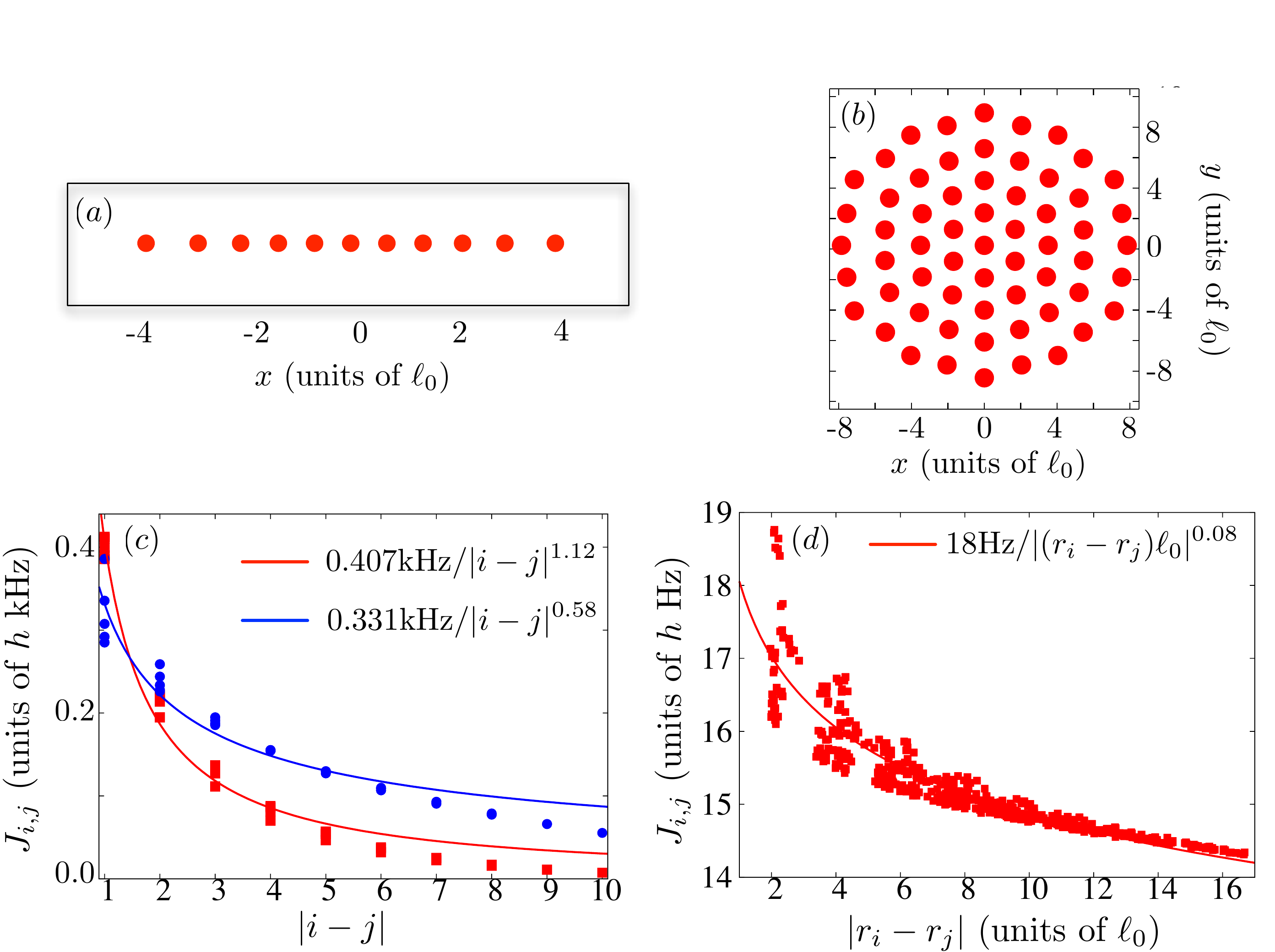}
\caption{(Color online) \emph{Equilibrium structures and effective spin-spin couplings for 1D and 2D ion traps.} (a) The equilibrium crystal structure for a 1D Paul trap with $\mathcal{N}_s=11$ ions and a trap anisotropy of $\omega_z/\omega_x=0.184$.  The spacing between ions is nonuniform, with the smallest spacing at the center of the chain.  (b) Equilibrium positions for $\mathcal{N}_s=61$ ions in a Penning trap, using the trap parameters of Ref.~\cite{bohnet2015quantum}.  Here, $\ell_0=(2 e^2/4\pi\varepsilon_0 M \omega_z^2)^{1/3}$.  The ions roughly form a triangular lattice.  (c) Spin-spin couplings Eq.~\eqref{eq:SSapprx} for the 1D trap with detuning $\delta/h=$80kHz above the center of mass mode (blue circles) and 150kHz above the center of mass (red squares), with the force adjusted so that the maximum value of $J_{i,j}$ is roughly 400Hz.  Best power law fits are shown with solid lines.  Significant deviations from translational invariance are seen, especially for the larger detuning.  (d) Spin-spin couplings Eq.~\eqref{eq:SSapprx} for the 2D trap with a detuning of $\delta/h=2$kHz above the center of mass and a spin-dependent force similar to Ref.~\cite{bohnet2015quantum}.  While there is some non-uniformity, the interactions are well-modeled by an all-to-all interaction, justifying the use of only the center of mass mode in the main text.
}
\label{fig:modesandJs}
\end{figure}

We numerically find the equilibrium positions of the ions following Ref.~\cite{james1998quantum} for the Paul trap and Ref.~\cite{PhysRevA.87.013422} for the Penning trap; typical equilibrium structures and effective spin-spin couplings are shown in Fig.~\ref{fig:modesandJs}.  The spin-spin interactions in the 2D array with detuning close to the center of mass, shown in Fig.~\ref{fig:modesandJs}(d), are well-approximated by a uniform, all-to-all interaction.  In contrast, the effective range of the interactions for typical parameters in linear Paul traps, modeled as a power law decay $J_{ij}\sim 1/|r_i-r_j|^{\alpha}$~\cite{britton2012engineered,PhysRevA.92.043405}, depends on the detuning, see Fig.~\ref{fig:modesandJs}(b)-(c).

Remarkably, an exact solution for the dynamics of Eq.~\eqref{eq:Hions} exists when starting from an uncorrelated product state of phonons and spins~\cite{Dylewsky}.  We will use this exact solution to numerically benchmark our methodologies in both the 2D case, where we keep only a single boson mode and use the swapping method described in Sec.~\ref{sec:MI}, and in the 1D case, where we keep all modes and use the methods of Sec.~\ref{sec:MII}.  In particular, we will compare the one- and two-point correlation functions
\begin{align}
\label{eq:exb}\langle \sigma^{a}_j\rangle=&\frac{1}{2}\prod_{\mu=1}^{\nb} e^{-2\left|\alpha_{\mu j}(t)\right|^2}\prod_{i\ne j}^{\ns}\cos\left(4J_{ij}\left(t\right)\right)\, ,
\end{align}
\begin{align}
\nonumber \langle \sigma^a_i \sigma^b_j\rangle=&\frac{1}{4}\prod_{\mu=1}^{\nb} e^{-2\left|a\alpha_{\mu i}(t)+b\alpha_{\mu j}(t)\right|^2}\\
&\times \prod_{k\ne i,j}^{\ns}\cos\left(4aJ_{i,k}(t)+4bJ_{j,k}(t)\right)\, ,
\end{align}
\begin{align}
\nonumber \langle \sigma^a_i \sigma^z_j\rangle=&\frac{ai}{2}\prod_{\mu=1}^{\nb} e^{-2\left|a\alpha_{\mu i}(t)\right|^2}\sin\left(4J_{i,j}(t)\right)\\
\label{eq:exe}&\times \prod_{k\ne i,j}^{\ns}\cos\left(4J_{i,k}(t)\right)\, .
\end{align}
where $a,b\in\{+,-\}$ and we have taken the initial state in which all phonons start in the vacuum state and all spins initial point along the $+x$ direction.  In general, the phonon temperature in trapped ion experiments is nonzero.  Our methods as presented here, as well as the exact solution, can be readily extended to any phonon pure state, and hence be used to generate thermal expectations by summing over Fock state expectations.  The effects of finite phonon temperature have been analyzed using the exact solution in Ref.~\cite{Dylewsky}.  A more efficient method for large numbers of phonons may be to use MPS methods specifically designed for finite-temperature systems~\cite{feiguin_white_05,white_09}.

In what follows, we also use an MPS ansatz which explicitly conserves the $\mathbb{Z}_2$ symmetry of Eq.~\eqref{eq:Hions} associated with the parity operators $\prod_{\mu}\left(-1\right)^{\hat{n}_{\mu}}\prod_j \hat{\sigma}^x_j$ and $\prod_j \hat{\sigma}^x_j$ for spin-boson and spin systems, respectively, as conservation of symmetries leads to significant computational gains in MPS algorithms~\cite{PhysRevA.82.050301,PhysRevB.83.115125}.  For the MPS calculations on pure spin models, we calculate the dynamics by fitting a long-range spin model with time-independent spin-spin couplings determined by Eq.~\eqref{eq:SSapprx} to an MPO, and then time-evolve using the local Runge-Kutta method of Ref.~\cite{Zaletel}.  As these spin-spin couplings are not translationally invariant, we use the non-uniform exponential MPO fitting method developed in Refs.~\cite{koller2016dynamics,Wallinprep}.

\begin{figure}
\centering
\includegraphics[width=0.75\columnwidth]{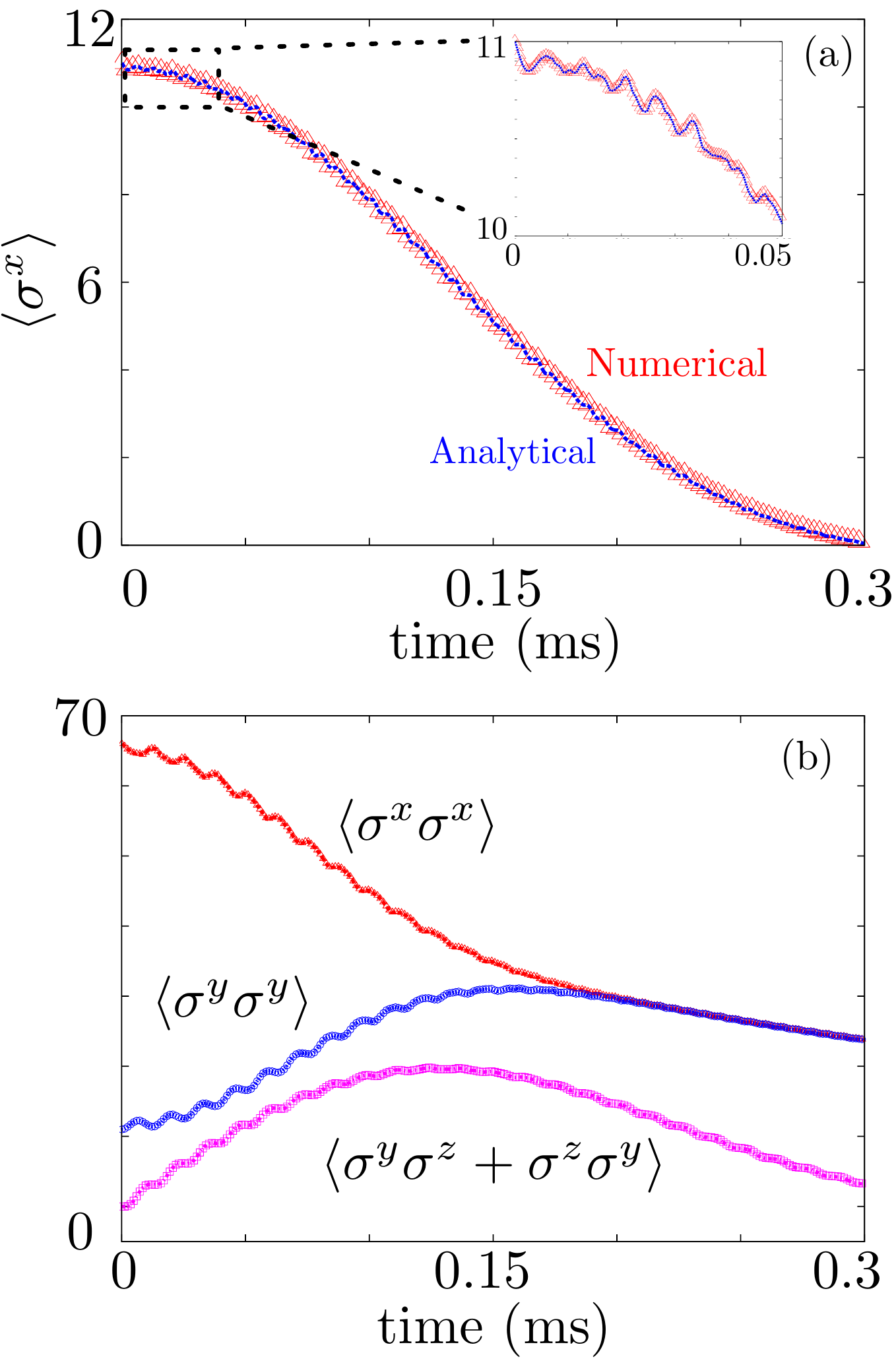}
\caption{(Color online) \emph{Comparison of numerical and analytically computed collective spin observables; multi-mode case.}  (a) Interaction-induced depolarization of the collective spin computed numerically (red symbols) and analytically (blue lines) for the $\delta/h=$80kHz detuned configuration of Fig.~\ref{fig:modesandJs}, showing excellent agreement.  The inset shows small oscillations due to spin-phonon entanglement, which are not captured within spin-only approaches.   (b) Collective spin correlations for the $\delta/h=$150kHz detuned configuration of Fig.~\ref{fig:modesandJs}.  The difference between the numerical values (symbols) and analytical results (lines) is not visible on the scale of the plot.
}
\label{fig:1DContrast}
\end{figure}

We begin our benchmarking by looking at the 1D case, in which the drive of the spin-dependent force is far detuned above the center of mass mode and so all phonon modes contribute to the dynamics.  In particular, we consider the detunings above the center of mass mode $\delta/h= 80$kHz and $\delta/h= 150$kHz, employed in Ref.~\cite{richerme2014non}, and which results in the effective spin-spin couplings shown in Fig.~\ref{fig:modesandJs}(c).  When the spin-dependent force is turned on, the effective Ising interactions will cause a coherent depolarization of the collective spin length $\langle {\sigma}^x\rangle=\sum_{i=1}^{\ns}\langle \hat{\sigma}^x_i\rangle$ concurrent with the development of collective spin-spin correlations $\langle \sigma^{a}\sigma^{b}\rangle=\sum_{i,j=1}^{\ns} \langle \hat{\sigma}^a_i\hat{\sigma}^b_j\rangle$.  In addition, spin-phonon entanglement resulting from spin-dependent displacements of the phonon modes will cause oscillations in collective spin observables.  Both effects are seen in Fig.~\ref{fig:1DContrast}, which compares the numerically computed collective spin expectations (red symbols) with the exact, analytic expressions Eq.~\eqref{eq:exb}-\eqref{eq:exe} (blue lines).  Panel (a) shows the depolarization of the collective spin for the $\delta/h=$80kHz detuning, with the inset showing the small amplitude oscillations due to dynamical phonons which are perfectly captured within our numerical approach.  Panel (b) shows the development of collective spin-spin correlations for the $\delta/h=$150kHz detuned situation.  Here, the numerical points are indistinguishable from the analytical curves on the scale of the plot, again demonstrating excellent agreement.

\begin{figure}
\centering
\includegraphics[width=0.75\columnwidth]{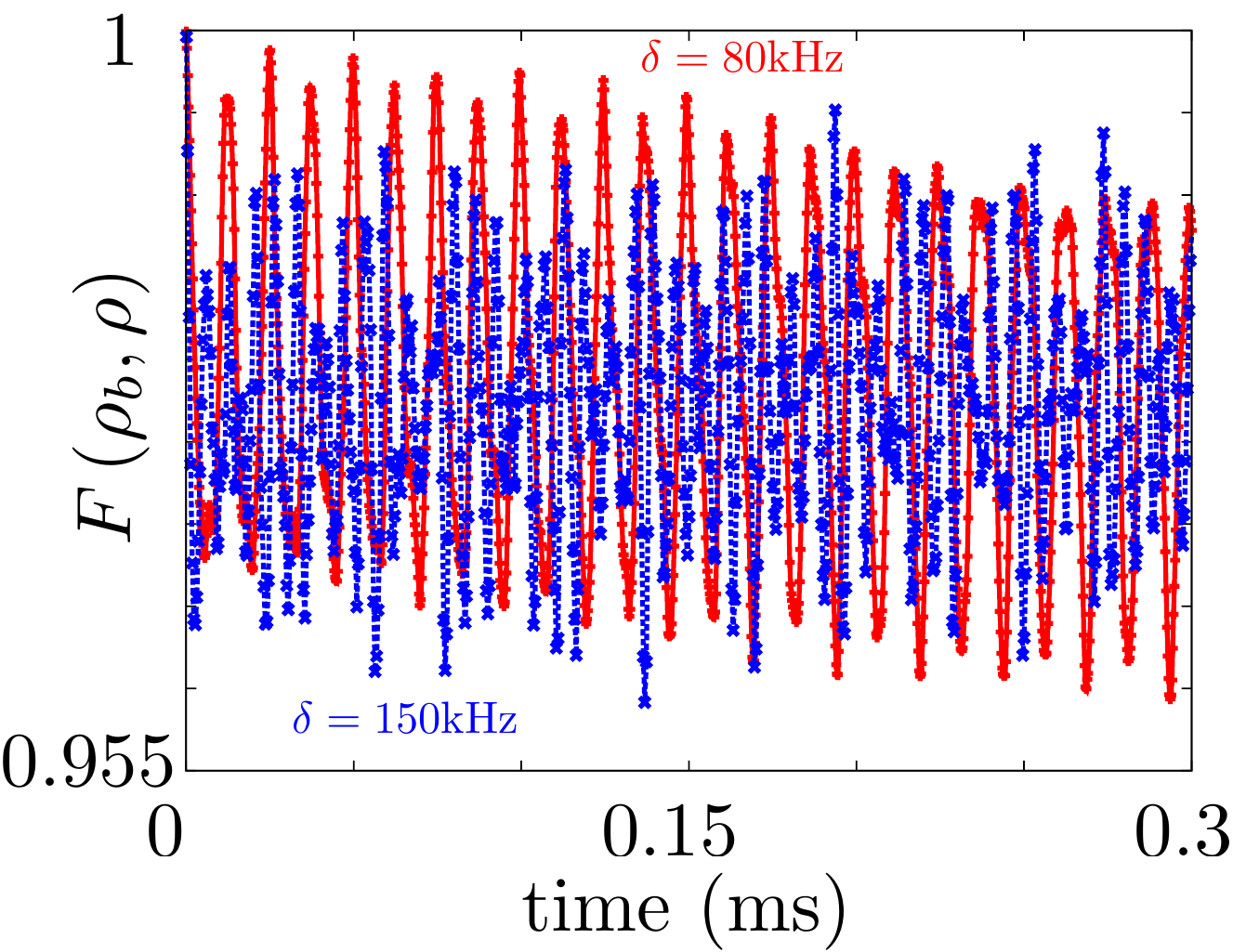}
\caption{(Color online) \emph{Fidelity of spin-boson dynamics with pure spin dynamics.}  The fidelity of the spin-boson dynamics with the bosons traced out with the pure spin dynamics, calculated using the methods of Sec.~\ref{sec:OaF}, is displayed for the $\delta/h=$80kHz detuning (red pluses) and the $\delta/h=$150kHz detuning (blue $\times$s) configurations of Fig.~\ref{fig:modesandJs}.  Rapid oscillations resulting from the development of spin-phonon entanglement with all phonon modes is visible, but the overall fidelity remains quite high.
}
\label{fig:1DFidelity}
\end{figure}

As mentioned above, the most stringent test of the fidelity of quantum simulation of a pure spin model can be obtained by measuring the fidelity of the spin reduced density matrix obtained from the wavefunction of the full spin-boson dynamics $|\psi\rangle$ by tracing out the bosons, $\rho_b=\mathrm{Tr}_{\mu}|\psi\rangle\langle \psi|$, with the pure spin density matrix $\rho$ obtained from evolution under the time-independent Ising model, Eq.~\eqref{eq:SSapprx}.  We compute this quantity using the methods of Sec.~\ref{sec:OaF}; the results are shown in Fig.~\ref{fig:1DFidelity} for the two detunings we consider.  We see rapid oscillations at many frequencies, corresponding to the generation of spin-phonon entanglement with all phonon modes.  However, throughout the course of reasonable experimental timescales, the fidelity remains quite high, $\gtrsim 95\%$.  We stress that the computation of this fidelity in the multimode case using the exact solution of the spin-boson dynamics~\cite{Dylewsky} scales poorly with the number of spins; in contrast, this quantity is easily calculated within our numerical framework

\begin{figure}
\centering
\includegraphics[width=0.75\columnwidth]{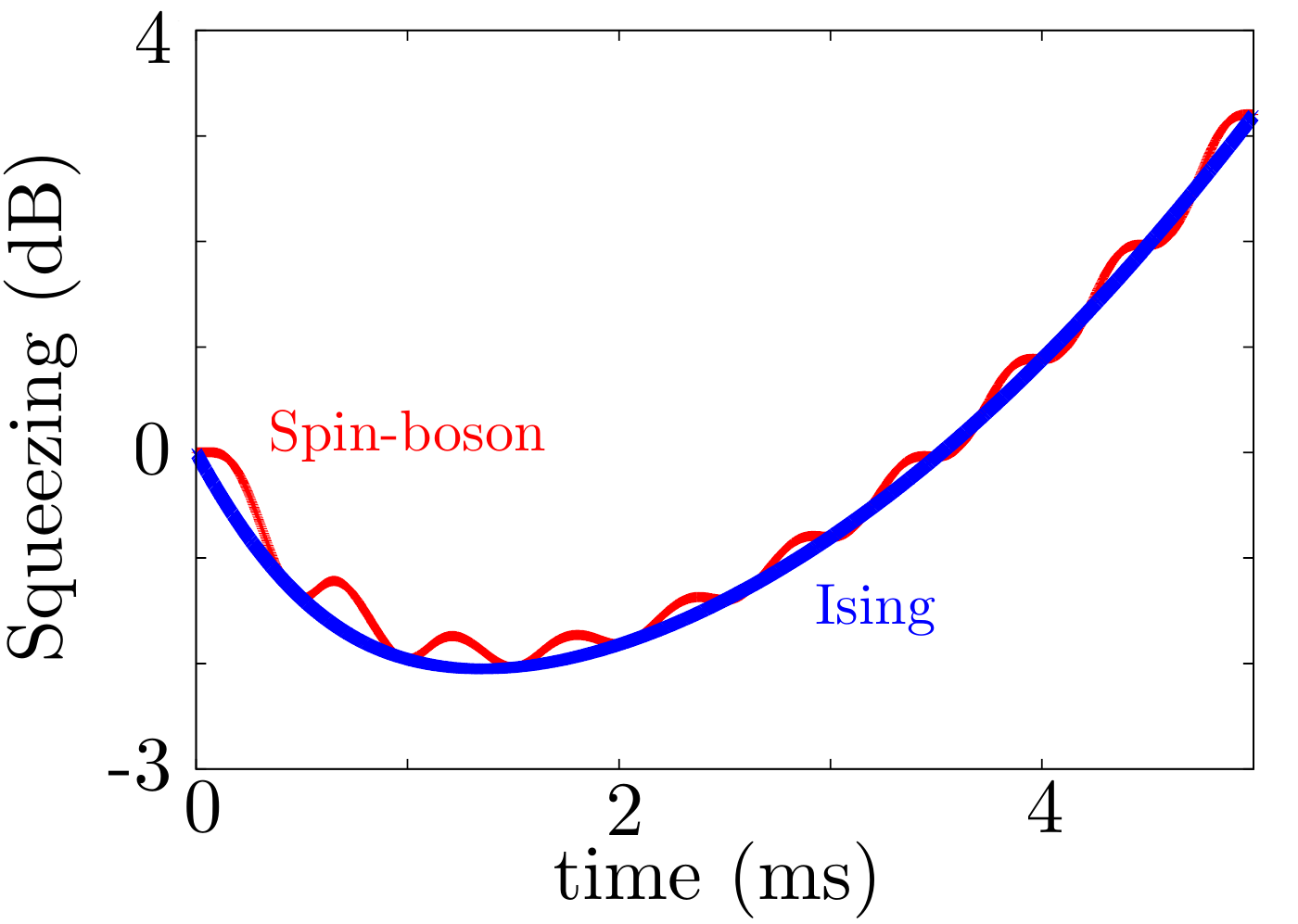}
\caption{(Color online) \emph{Squeezing dynamics; single-mode case.}  Dynamics of the Ramsey squeezing parameter in dB for the coupled spin-boson system (oscillating red points) and the pure spin system (blue points) evolving under Eq.~\eqref{eq:SSapprx}, with parameters from Fig.~\ref{fig:modesandJs}(d).  Coupling to phonons as well as time dependence of the Ising spin-spin couplings (see Eq.~\eqref{eq:SpinSpinTD}) cause modulation of the squeezing in the spin-boson dynamics.  The spin-boson and pure spin approaches agree at the decoupling points, here given by integer multiples of $0.5$ms.
}
\label{fig:2DSqueezing}
\end{figure}

We now turn to the 2D case, in which only the center of mass mode participates in the dynamics and we use the swapping method (Sec.~\ref{sec:MI}) to compute the dynamics.  We use the parameters of Fig.~\ref{fig:modesandJs}, which are similar to recent experiments~\cite{bohnet2015quantum}.  A key observable in these experiments is the Ramsey squeezing parameter, defined as~\cite{PhysRevA.47.5138}
\begin{align}
\label{eq:Squeezdef} \xi&=\min_{\theta} \frac{\sqrt{\ns \Delta S_{\theta}^2}}{|\langle {S}_x\rangle|}\, ,
\end{align}
with $\Delta S_{\theta}^2=\langle \hat{S}_{\theta}^2\rangle-\langle \hat{S}_{\theta}\rangle^2$ and $\hat{S}_{\theta}=\cos\theta\hat{S}^z+\sin\theta\hat{S}^y$.  The squeezing parameter is of interest because the condition $\xi^2<1$ witnesses entanglement in the system~\cite{PhysRevLett.86.4431}, and $\xi^2$ also quantifies the signal to noise enhancement of the spin state for performing Ramsey spectroscopy compared to an uncorrelated spin state~\cite{PhysRevA.46.R6797}.  Fig.~\ref{fig:2DSqueezing} displays the squeezing as a function of time starting from all spins aligned along $x$, and compares this with the results of simulating the time-independent Ising model Eq.~\eqref{eq:SSapprx}.  While the pure spin approach predicts a smooth buildup of squeezing to a maximal point followed by monotonic destruction of squeezing, the spin-boson dynamics displays oscillations resulting from the development of spin-phonon entanglement, which tends to antisqueeze the spin distribution, as well as from the time dependence of the spin-spin couplings $\tilde{J}_{ij}\left(t\right)$( see Eq.~\eqref{eq:SpinSpinTD}).  However, the two results agree at the decoupling times $t_d=2\pi \hbar n/\delta$, $n$ an integer, where the spin-phonon entanglement vanishes.  While we do not show it here, we again find excellent agreement between our computed values and the exact solution, validating the swapping approach to dynamics.

\begin{figure}
\centering
\includegraphics[width=0.99\columnwidth]{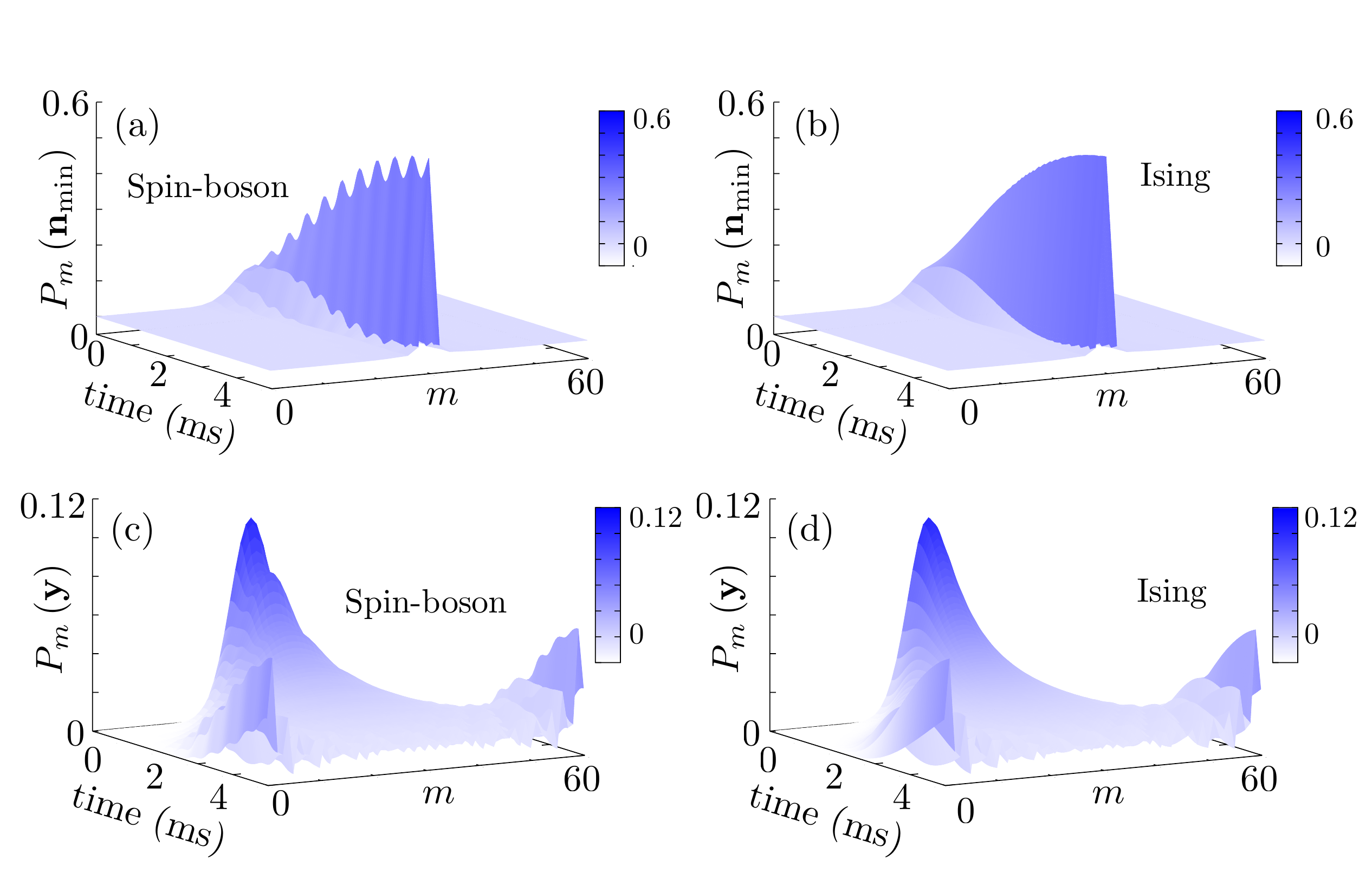}
\caption{(Color online) \emph{Full counting statistics of spin-boson and pure spin states.} The full counting statistics, obtained using the methods of Sec.~\ref{sec:OaF}, are displayed for the squeezed quadrature minimizing the spin variance (upper panels) as well as the antisqueezed quadrature along $\hat{S}^y$ (lower panels) for the parameters of Fig.~\ref{fig:modesandJs}(d).  The left panels display the full coupled spin-boson dynamics, and the right panels correspond to spins evolving under a time-independent Ising model.  Periodic modulations from spin-phonon entanglement broaden the squeezed quadrature and can either broaden or narrow features in the antisqueezed quadrature.  Generally, the antisqueezed quadrature is less sensitive to phonon coupling than the squeezed quadrature.
}
\label{fig:2DFCS}
\end{figure}

The modification of spin observables due to spin-phonon entanglement is strikingly seen by comparing the full counting statistics (FCS) of collective spin observables between the spin-boson and pure spin approaches.  The computation of these quantities from the analytical solution scale polynomially in the number of spins in the case that only the center of mass mode participates in the dynamics~\cite{bohnet2015quantum}, but also requires high-precision arithmetic to avoid catastrophic cancellation.  In the general case, the computation of these quantities scales exponentially in $\ns$.  However, as discussed in Sec.~\ref{sec:OaF}, the computation of the FCS is particularly straightforward within our MPS-based approach.  Fig.~\ref{fig:2DFCS} displays the FCS along the antisqueezed quadrature $\mathbf{y}$ as well as the squeezed quadrature $\mathbf{n}_{\mathrm{min}}$ such that $\mathbf{S}\cdot \mathbf{n}_{\mathrm{min}}$ minimizes the spin variance $\Delta S^2$.  The upper panels (a)-(b) display the squeezed quadrature, which show the development of a narrow feature which is useful for sub-shot-noise spectroscopy, with panel (a) corresponding to the coupled spin-boson dynamics and panel (b) to the pure spin dynamics.  The pure spin dynamics shows a monotonic development of this narrow feature, while this feature is periodically broadened and its height modulated in the spin-boson dynamics due to spin-phonon entanglement.  The antisqueezed quadratures, shown in panels (c)-(d), initially broaden with respect to the initial uncorrelated state before developing fine-scale features whose usefulness for metrology are not witnessed by spin squeezing but can be captured via other measures, such as the quantum Fisher information~\cite{PhysRevLett.102.100401,bohnet2015quantum}.  A comparison between the pure spin and spin-boson calculations again shows periodic differences driven by spin-phonon entanglement, but the differences are more slight compared to the squeezed quadrature.

\section{Including decoherence: Quantum Trajectories}
\label{sec:QT}

An important component of many AMO experiments is losses and other forms of dissipation, which are usually thought to degrade coherence and destroy quantum correlations, but can also lead to the production of entangled steady states~\cite{PhysRevLett.107.080503,PhysRevLett.109.230501,lin2013dissipative}.  A major source of decoherence in trapped ion systems arises from scattering of the light used to create the spin-dependent force~\cite{PhysRevLett.105.200401}.  A full description of the system including losses is found by solving the master equation
\begin{align}
\hbar \frac{d}{dt}\hat{\rho}&=-i\left[\hat{H},\hat{\rho}\right]-\hat{\mathcal{L}}\hat{\rho}\, ,
\end{align}
where $\hat{\rho}$ is the system density matrix and the action of the Lindbladian superoperator accounting for spontaneous emission from the Raman beams is given as
\begin{align}
\nonumber \hat{\mathcal{L}}\hat{\rho}=&\frac{\Gamma_{\mathrm{ud}}}{2} \sum_j \left(\{\hat{\sigma}^+_j\hat{\sigma}^{-}_j, \rho\}-2\hat{\sigma}^-_j\rho\hat{\sigma}^+_j\right)\\
\nonumber &+\frac{\Gamma_{\mathrm{du}}}{2} \sum_j \left(\{\hat{\sigma}^-_j\hat{\sigma}^{+}_j, \rho\}-2\hat{\sigma}^+_j\rho\hat{\sigma}^-_j\right)\\
&+\frac{\Gamma_{\mathrm{el}}}{4} \sum_j \left(\rho-\hat{\sigma}^z_j\rho\hat{\sigma}^z_j\right)\, .
\end{align}
Here, the terms proportional to $\Gamma_{\mathrm{du}}$ and $\Gamma_{\mathrm{ud}}$ correspond to spontaneous excitation and de-excitation, respectively, from Raman scattering, and the term proportional to $\Gamma_{\mathrm{el}}$ represents elastic dephasing.

It is possible to apply MPS-based approaches to directly simulate master equations by replacing states by operators (``superkets") and operators by superoperators, in which the local dimension of superket $i$ is $d_i^2$, the square of the physical on-site Hilbert space~\cite{PhysRevLett.93.207205,PhysRevLett.93.207204}.  Here, we instead use the method of quantum trajectories, in its first-order incarnation~\cite{RevModPhys.70.101,doi:10.1080/00018732.2014.933502}.  To do so, it is useful to re-write the master equation as
\begin{align}
\hbar \frac{d}{dt}\hat{\rho}&=-i\left(\hat{H}_{\mathrm{eff}}\hat{\rho}-\hat{\rho}\hat{H}_{\mathrm{eff}}^{\dagger}\right)+\hat{\mathcal{R}}\hat{\rho}\, ,
\end{align}
where the non-Hermitian effective Hamiltonian is
\begin{align}
\hat{H}_{\mathrm{eff}}&=\hat{H}-\frac{i}{2}\sum_{\mu}\hat{J}^{\dagger}_{\mu} \hat{J}_{\mu}
\end{align}
and the recycling term is
\begin{align}
\mathcal{R}\rho&=\sum_{\mu} \hat{J}_{\mu}\rho\hat{J}_{\mu}^{\dagger}\, .
\end{align}
In these expressions, the ``jump operators" $\{\hat{J}_{\mu}\}$ are a basis in which the Lindbladian may be written as
\begin{align}
\hat{\mathcal{L}}\rho&=-\frac{1}{2}\sum_{\mu} \left[\{\hat{J}_{\mu}^{\dagger}\hat{J}_{\mu},\rho\}-2\hat{J}_{\mu}\rho \hat{J}_{\mu}^{\dagger}\right]\, .
\end{align}
In the present case, we take the basis of jump operators to be the $3\ns$ operators $\{\sqrt{\Gamma_{\mathrm{ud}}}\hat{\sigma}^-_j,\sqrt{\Gamma_{\mathrm{du}}}\hat{\sigma}^+_j,{\sqrt{\Gamma_{\mathrm{el}}}}\hat{\sigma}^z_j/{2}\}$ for $j=1\dots \ns$.  Now, at each time step, we evolve $|\psi\left(t\right)\rangle$ under $\hat{H}_{\mathrm{eff}}$ to obtain $|\tilde{\psi}\left(t+\delta t\right)\rangle$.  Because the effective Hamiltonian is non-Hermitian, the norm of this state, call it $\left(1-\delta p\right)$, will be less than 1.  At first order in $\delta t$, $\delta p$ is comprised of a sum of terms $\delta p=\sum_{\mu}\delta p_{\mu}$, where
\begin{align}
\delta p_{\mu'}&=\langle \psi\left(t\right)|\hat{{J}}_{\mu}^{\dagger}\hat{{J}}_{\mu}|\psi\left(t\right)\rangle\, .
\end{align}
We now stochastically choose whether to use this state at the next time step, or undergo a ``quantum jump" by applying the recycling term.  Namely, we choose a random number $r\in\left[0,1\right]$, and select the state at the next time step to be
\begin{align}
\nonumber |\psi\left(t+\delta t\right)\rangle&=\left\{\begin{array}{c} |\tilde{\psi}\left(t+\delta t\right)\rangle\;\; \;\mbox{with probability }1-\delta p\, ,\\
\hat{\mathcal{J}}_{\mu}|\psi\left(t\right)\rangle \;\;\;  \mbox{with probability }\delta p_{\mu}\end{array} \right. \, ,
\end{align}
and then renormalizing the state.  The equally weighted average of the expectation of an operator $\langle \hat{O}\rangle$ taken with ${N}_{\mathrm{samp}}$ such stochastically generated trajectories converges to the dynamics of this same observable generated by the master equation with an error $\varepsilon_{\hat{O}}$ that is asymptotically statistical, $\varepsilon_{\hat{O}}\sim 1/\sqrt{N_{\mathrm{samp}}}$.  In closing, we note that the quantum trajectories method may be trivially parallelized.

\section{Application: Trapped ion spin-boson simulators in the presence of molecular impurities}
\label{sec:Impure}

For a pure spin Ising model (i.e., without dynamical phonons), the summation over all quantum trajectories can be performed analytically, leading to an exact solution of arbitrary Ising models in the presence of decoherence~\cite{PhysRevA.87.042101, 1367-2630-15-11-113008}.  When the dynamical motion of the phonons is included, however, no closed-form summation over quantum trajectories exists, and the dynamics must be obtained numerically.  In what follows, we will consider the dynamics in the presence of decoherence as parameterized above in the 2D Penning trap where only the center of mass mode is relevant.  In principle, these dynamics can be calculated efficiently, even in the presence of decoherence, for a spatially uniform initial state by restricting the Hilbert space to the space of permutationally symmetric density matrices~\cite{Carmichael,hartmann2012generalized, PhysRevB.91.035306}.  In order to exemplify the power and generality of our approach compared to these other methods, we thus consider a case in which the center of mass mode is no longer spatially uniform.  In particular, we simulate the case of an ion crystal in which some atomic ions have converted to heavier mass molecular ions through collisions with background gas, focusing on a crystal of Be$^+$ ions with BeH$^+$ impurities as a particular example~\cite{PhysRevA.91.011401}.  

\begin{figure}
\centering
\includegraphics[width=0.7\columnwidth]{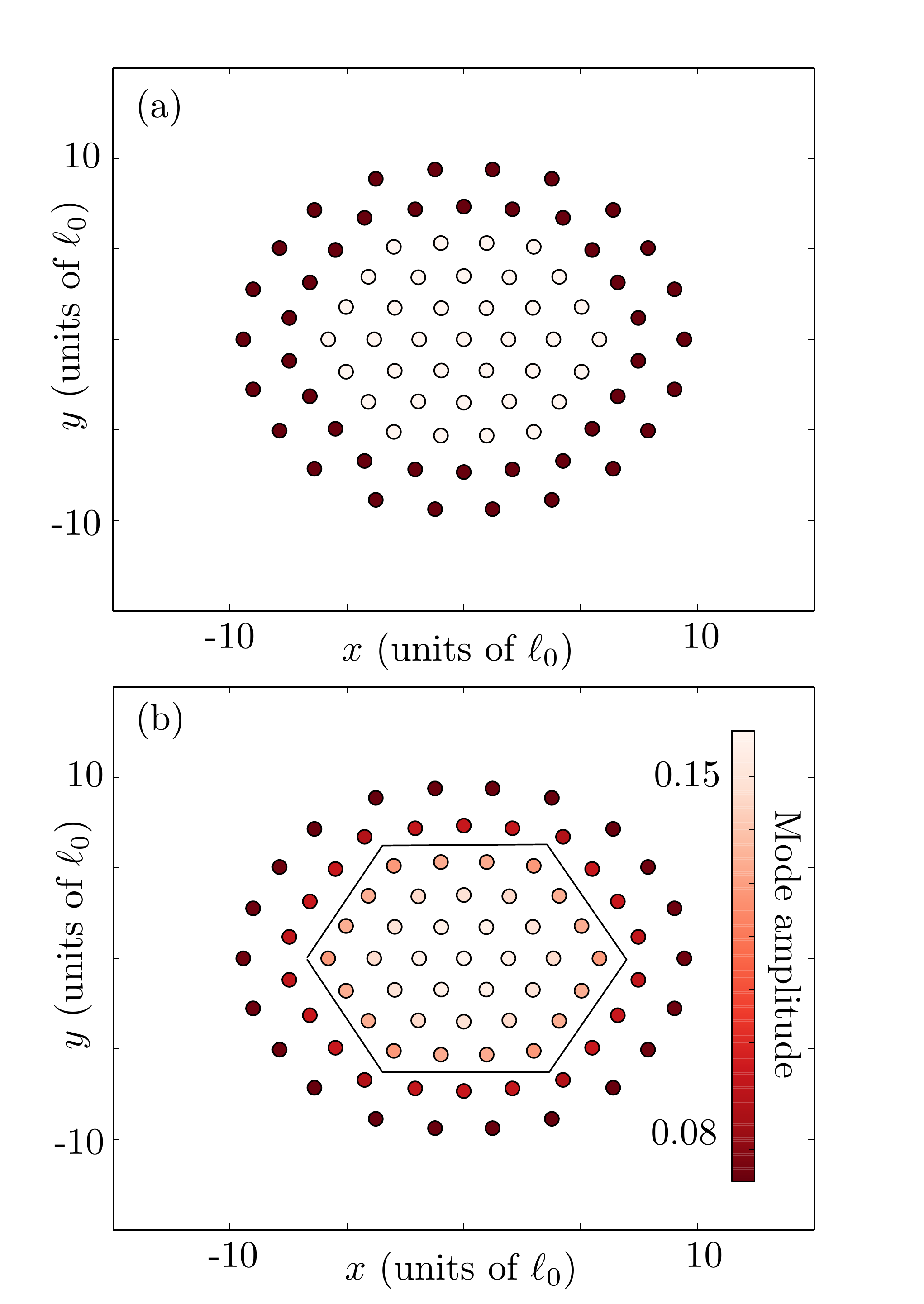}
\caption{(Color online) \emph{Equilibrium structures and center of mass mode of an ion crystal with impurities.} (a) Equilibrium crystal structure for 37 atomic Be$^{+}$ ions (empty circles) and 44 BeH$^+$ molecular ions (filled circles) in a Penning trap.  The heavier molecular ions move to the outside of the crystal due to centrifugal forces in the rotating frame of the ion crystal.  (b) Amplitude of the highest frequency axial normal mode, analogous to the center of mass mode in a crystal without impurities.  The molecular ions have a reduced amplitude due to their larger mass, and the amplitudes of the atomic ions (inside boxed region) are spatially non-uniform, which in turn makes the spin-phonon coupling spatially inhomogeneous.
}
\label{fig:DC}
\end{figure}

Impurity molecular ions have a heavier mass than the atomic ions, and so move to the edge of the rotating crystal due to centrifugal forces.  We compute the equilibrium positions and axial normal modes of the ion crystal in the presence of these impurities using the methods of Ref.~\cite{PhysRevA.88.043434}; results for a crystal of 37 Be$^{+}$ ions and 44 BeH$^+$ impurities are shown in Fig.~\ref{fig:DC}.  Panel (a) shows the equilibrium crystal structure for trapping parameters similar to current experiments~\cite{bohnet2015quantum}, with filled (empty) circles denoting molecular impurities (atomic ions).  The highest frequency axial mode, analogous to the center of mass mode in a crystal with no impurities, is shown in in panel (b).  The impurity ions have a greatly reduced amplitude of oscillation due to their heavier mass, and, more importantly for our purposes, the amplitudes of the center of mass mode on the atomic ions are non-uniform.  The spread in amplitudes is roughly 10\% of the mean for the particular parameters we consider.  Only the atomic ions respond to the light-induced spin-dependent force that generate effective Ising spin-spin interactions, and so the effective number of spins in the quantum spin simulator is the number of atomic ions: 37 in our example case.  Still, the molecular ions influence the spin-phonon and spin-spin dynamics through their modification of the center of mass mode amplitudes on the atomic ions.

\begin{figure}
\centering
\includegraphics[width=0.7\columnwidth]{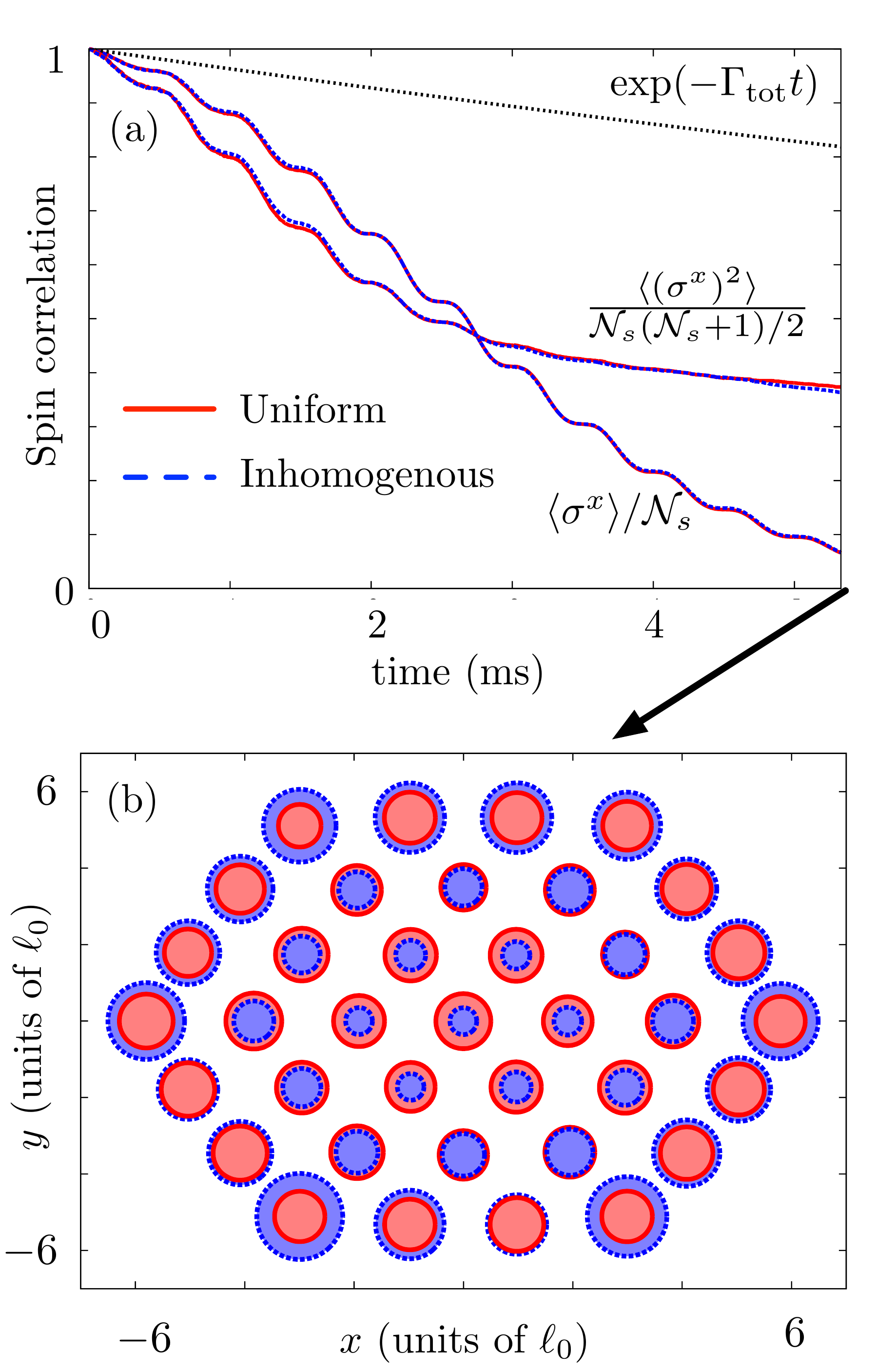}
\caption{(Color online) \emph{Spin correlations in an ion crystal with impurities and decoherence.} (a) Normalized collective magnetization $\langle \sigma^x\rangle$ and its square $\langle (\sigma^x)^2\rangle$  computed for the spatially inhomogeneous mode pictured in Fig.~\ref{fig:DC}(b) (blue dashed) and its closest uniform approximation (red solid).  The black dashed line is the mean-field prediction of the decay of magnetization due to decoherence.  (b) Spatially resolved magnetization $\langle \sigma^x(x,y)\rangle$ at the final time shown in (a), with circle size indicating magnitude.  Blue, dashed border (red, solid border) circles are the inhomogeneous mode (uniform approximation).  Note that all red circles have the same diameter, and only the blue circles are inhomogeneous.  The spread in magnetization values is roughly twice the mean for the inhomogeneous mode.
}
\label{fig:DCDyn}
\end{figure}

Our calculations of the spin dynamics for a spatially non-uniform center of mass mode with decoherence are shown in Fig.~\ref{fig:DCDyn}.  We take decoherence rates of $\Gamma_{\mathrm{el}}=60$s$^{-1}$, $\Gamma_{\mathrm{ud}}=9$s$^{-1}$, and $\Gamma_{\mathrm{ud}}=6$s$^{-1}$, comparable to current experiments, and average over $\sim200$ trajectories.  Panel (a) shows collective spin correlations such as the normalized total magnetization $\langle \sigma^x\rangle $ and its square $\langle (\sigma^x)^2\rangle $.  The black dashed line is the mean-field prediction of the decay of magnetization, which occurs solely due to decoherence at a rate $\Gamma_{\mathrm{tot}}=(\Gamma_{\mathrm{el}}+\Gamma_{\mathrm{ud}}+\Gamma_{\mathrm{ud}})/2$.  Strong deviations from this curve indicate the decay of magnetization due to the build-up of higher-order spin correlations.  The blue dashed lines for the spin correlations use the actual inhomogeneous analog of the center of mass mode shown in Fig.~\ref{fig:DC}(b).  In contrast, the solid red lines use its nearest uniform approximation, in which the amplitude of the mode at each position is replaced by the average of the amplitude vector.  Almost no difference between the exact and approximate predictions can be seen in the decay of the magnetization, whereas a slight difference can be seen in the squared collective magnetization at late times.  In contrast, the effect of the mode inhomogeneity is clearly seen in the spatially resolved magnetization shown in Fig.~\ref{fig:DCDyn}(b).  Here, the size of the circles are proportional to $\langle \sigma^x(x,y)\rangle$ and the red solid (blue dashed) circles are for the uniform approximation and the true inhomogeneous mode, respectively.  Our results demonstrate that, even in the presence of a relatively large number of impurity ions, the inhomogeneities in the analog of the center of mass mode have only a slight impact on low-order collective spin correlations over experimentally relevant timescales, whereas inhomogeneities in the spatial distribution of the magnetization can be quite sizable.

\section{Conclusions and outlook}
\label{sec:Concl}
We have developed a generic approach to the out-of equilibrium dynamics of spins globally coupled to bosonic modes, which encompasses many paradigmatic models from all areas of quantum science, based on matrix product states (MPSs).  In contrast to existing methods, our framework applies for any spatial and operator dependence of the spin-boson coupling, and places no restrictions on relative energy scales.  In the regime of many fewer bosons than spins, $\mathcal{N}_s\gg \mathcal{N}_b$, an efficient approach is to use a Trotter-Suzuki decomposition of the spin-boson coupling propagator and dynamically deform the MPS topology using swap gates, resulting in a method with rigorously controlled error.  Alternatively, for equal numbers of spins and bosons, $\mathcal{N}_s=\mathcal{N}_b$, we propose to perform direct time evolution using an exact matrix product operator (MPO) representation of the spin-boson Hamiltonian with long-ranged couplings, and provide techniques to safeguard against variational metastability.  This exact MPO representation of the Hamiltonian can also be employed to find eigenstates and static properties of spin-boson systems.

Using our newly developed approach, we calculated the dynamics of 1D and 2D trapped ion quantum simulators in experimentally relevant parameter regimes, fully accounting for the dynamics of the phonons that mediate effective spin dynamics between ions.  We first benchmarked our numerical results against a recent exact analytical solution for a particular spin-phonon coupling, finding excellent agreement.  Beyond comparing low-order correlation functions which are straightforward and efficient to compute with the exact solution, we also demonstrated the power of our approach for computing detailed properties of the quantum state which are difficult to obtain through other means, such as the full counting statistics of collective spin observables and the fidelity of the spin-phonon system as a pure spin quantum simulator.  

We showed that Markovian decoherence, which is an important component of trapped ions and many other AMO platforms, can be incorporated within our approach by solving the associated master equation using the method of quantum trajectories.  Within this framework, we simulated the dynamics of a 2D ion crystal in the presence of realistic decoherence.  In order to evince the power of our approach beyond those that require spatial uniformity of the spin-boson coupling, we treated the case of a non-uniform spin-boson coupling, physically motivated by the presence of impurity molecular ions in an atomic ion quantum spin simulator.

While our results show that MPSs are a powerful framework for studying spin-boson models, there are certain regimes which remain difficult or impossible to simulate efficiently with MPSs.  In particular, 2D systems with many boson modes and finite-range interactions are challenging, and it is known that the scaling of MPSs for finding equilibrium states in 2D is exponential in the system size.  Even in one dimension, simulations in which spins are strongly entangled with many bosons in a spatially inhomogeneous fashion are difficult, and particularly subject to variational metastability.  An additional source of difficulty from a numerical point of view is that there is often a large separation of timescales between the dynamics of direct spin-boson coupling and boson-mediated spin-spin interactions, with the former limiting the size of the time step.  Here, applying multiscale methods~\cite{KevorkianJK1996} is a promising avenue for improving numerical performance.

Myriad future research directions are available based on the framework developed in this paper.  For one, our approach is able to treat several competing interaction scales, as well as decoherence, on the same footing, and so provide a rigorous means for justifying effective models of driven spin-boson systems in various parameter regimes.  This is especially important for trapped ion quantum spin simulators in effective transverse fields, where the presence of non-commuting interactions with comparable energy scales can fundamentally change the effective spin physics in ways not captured by a perturbative approach~\cite{Wallinprep2}.  Additionally, while we focused on applications to trapped ion systems, our techniques can be applied to many other platforms, including quantum optics and optomechanics, and provide a scalable means for studying strongly correlated physics in these diverse systems.  Our framework also applies to higher-dimensional spin representations, such as spin-1, which can also be realized with trapped ion systems~\cite{PhysRevLett.112.040503,PhysRevX.5.021026} and can shed light on fundamental questions such as the nature of topological phases with long-range interactions~\cite{PhysRevB.93.041102}.  A generalization from discrete boson spectra to continuous boson spectra is possible, given a single global coupling operator between spins and bosons, using techniques based on orthogonal polynomials~\cite{PhysRevLett.105.050404,Chin.Plenio_10} while keeping well-defined error~\cite{PhysRevLett.115.130401}.  This can lead to new insights into quantum phases of correlated spins which are coupled to an external bath.

\section{Acknowledgements}

We would like to acknowledge useful discussions with Justin Bohnet, John Bollinger, Johannes Schachenmayer, Zhexuan Gong, and Phil Richerme, and support from NSF-PHY 1521080, JILA-NSF-PFC-1125844, ARO, MURI-AFOSR and AFOSR.  MLW thanks the NRC postdoctoral program for support.

\appendix
\section{Improving the convergence of long-range time evolution}
\label{sec:VM}

In this appendix, we discuss additional safeguards against variational metastability when using the long-range time evolution methods of Sec.~\ref{sec:MII}.  Time evolution proceeds by optimization of the functional $\min_{|\phi\rangle}||\phi\rangle-\hat{U}|\psi\rangle|^2$, with $\hat{U}$ an MPO representation of the propagator over a small time step $\delta t$~\cite{Zaletel} and $|\psi\rangle$ an MPS representation of the state at time $t$, over all MPSs $|\phi\rangle$ with fixed resources.  The optimal MPS $|\phi\rangle$ then becomes the new state at time $(t+\delta t)$.  The global optimization over the entire MPS cannot be performed efficiently, and so optimization is performed over the tensors of two neighboring sites in the MPS representation (see Eq.~\eqref{eq:MPSdef}) with all other tensors held fixed~\cite{Schollwoeck}.  A complete optimization cycle over all neighboring pairs of sites is referred to as a sweep.  In this local optimization scheme, the main cause of metastability is that entanglement is only built up between neighboring sites in a single step, and longer-ranged entanglement must be built up by repeated sweeping.  

In some cases, the algorithm is not able to properly build up the appropriate entanglement structure for a long-ranged Hamiltonian, especially when starting from an unentangled state of spins and bosons.  Our idea to improve the convergence is to also endow our initial variational guess $|\phi\rangle$ with long-range coherence between spins and bosons which are far-separated in the MPS representation.  We do so by generating a state $|\phi'\rangle$ which contains long-range coherence between all spins and all bosons and mixing in a small amount $\alpha$ of this state to the optimal state obtained from local variational minimization, with the mixing parameter $\alpha$ taken to zero towards the end of the sweeping optimization.  More precisely, the mixing is done by projecting the state $|\phi'\rangle$ onto the orthonormal basis formed by holding all tensors except two in the MPS representation fixed, and then adding this projected two-site wavefunction directly to the wavefunction obtained from the optimization procedure.  

This approach is similar in spirit to  the modified single-site DMRG algorithm by White~\cite{PhysRevB.72.180403}, which also builds fluctuations on top of the local variational optimum to accelerate convergence.  While a detailed study of the optimal states $|\phi'\rangle$ and mixing parameters $\alpha$ is outside of the scope of this work, we find that choosing $|\phi'\rangle$ to be the ground state of $\hat{H}$ and $\alpha\sim \mathcal{O}\left(0.1\right)$ to perform well.  Note that the ground state can be efficiently found using ordinary two-site DMRG with the MPO representation of Fig.~\ref{fig:IIdia}, as entanglement which is directly built up between neighboring spins and bosons propagates to farther separated spins and bosons by repeated sweeping.  We also note that an accurate representation of the ground state is not essential--only non-vanishing long-range coherence is required--and so relatively coarse tolerances can be used in obtaining the ground state.

\bibliography{references}

\end{document}